\documentclass[11pt,a4paper]{article}
\pdfoutput=1
\usepackage{jcappub}
\usepackage{bm}
\usepackage{empheq}
\usepackage[utf8]{inputenc}
\usepackage[toc,page]{appendix}

\newcommand{\be}{\begin{equation}}
\newcommand{\ee}{\end{equation}}
\newcommand{\bea}{\begin{eqnarray}}
\newcommand{\eea}{\end{eqnarray}}

\newcommand{\vect}[1]{\bm{\mathrm{{#1}}}}

\graphicspath{{./}}
\notoc

\usepackage[T1]{fontenc} 

\usepackage{braket}

\numberwithin{equation}{subsection}

\newcommand\ben{\begin{enumerate}}
\newcommand\een{\end{enumerate}}
\newcommand\bal{\begin{align*}}
\newcommand\eal{\end{align*}}
\newcommand\bi{\begin{itemize}}
\newcommand\ei{\end{itemize}}

\def\id{\protect{{1 \kern-.28em {\rm l}}}}

\newcommand{\gae}{\lower 2pt \hbox{$\, \buildrel {\scriptstyle >}\over {\scriptstyle
\sim}\,$}}
\newcommand{\lae}{\lower 2pt \hbox{$\, \buildrel {\scriptstyle <}\over {\scriptstyle
\sim}\,$}}

\usepackage{graphicx}
\usepackage{caption}
\usepackage{subcaption}

\begin{document}

\title{Numerically evaluating the bispectrum in curved field-space \\  {\LARGE \it  -- with PyTransport 2.0}}

\author{John W. Ronayne}
\author{and David J. Mulryne}
\affiliation{
School of Physics and Astronomy, Queen Mary
University of London, Mile End Road, London, E1 4NS, UK}

\emailAdd{j.ronayne@qmul.ac.uk; d.mulryne@qmul.ac.uk}
---------------------------------------------------------------------------------
\abstract{
\noindent We extend the transport framework for numerically evaluating the power spectrum and bispectrum in multi-field inflation to the case of a curved field-space metric. This method naturally accounts for all sub- and super-horizon tree level effects, including those induced by the curvature of the field-space. We present an open source implementation of our equations in an extension of the publicly available \texttt{PyTransport} code. Finally we illustrate how our technique is applied to examples of inflationary models with a non-trivial field-space metric.
\vspace{80mm}
}

\maketitle

\section{Introduction} 
Recently a convenient framework was developed by Dias {\it et al.} \cite{Dias:2016rjq} to numerically calculate the 
primordial power spectrum and bispectrum of the curvature perturbation, $\zeta$, 
produced by inflation with an arbitrary number of fields  (see also Ref.~\cite{Mulryne:2013uka,Anderson:2012em,Seery:2012vj,Mulryne:2010rp,Mulryne:2009kh,Dias:2011xy} for earlier related works\footnote{Early work on the 
transport approach considered only the super-horizon evolution of perturbations, however it was shown in Ref.~\cite{Mulryne:2013uka} that the approach could be extended to sub-horizon scales, and this work was used as a basis for   Ref.~\cite{Dias:2016rjq}.}). 
The essence of the approach is to set up coupled ordinary differential equations (ODEs) for the correlations 
of the inflationary fields' fluctuations. These correlations can then be related to the correlations of the curvature perturbation.
The framework accounts for all tree level effects on sub- and super-horizon scales, and is 
referred to as the ``transport approach''
to inflationary perturbations.

The work of Dias {\it et al.} \cite{Dias:2016rjq} presented a rather general framework,  
but specific equations were only given for inflation driven by multiple canonical scalar fields
with Euclidean field space metric, and only this case 
was implemented in two numerical codes \cite{Seery:2016lko, Mulryne:2016mzv} which accompanied the paper. 
The primary goal of the present work, therefore,
is to present explicit equations for the more general case where the 
field-space metric of the multi-field system is non-Euclidean.
At the level of the power spectrum 
the transport method has already been extended to this case, and a code 
released in the form 
of a Mathematica worksheet, \texttt{mTransport}, by Dias, Frazer and Seery~\cite{Dias:2015rca}. Here 
we extend this work to the bispectrum, presenting all the elements needed to 
implement the framework of Ref.~\cite{Dias:2016rjq} in this more general setting. An online resource for the 
transport method and the various codes (including the new contributions discussed below) 
is available at \href{https://transportmethod.com}{transportmethod.com}.

The two numerical packages which accompanied Ref.~\cite{Dias:2016rjq} 
represent the first publicly available tools developed 
to calculate the bispectrum numerically in
a multi-field model.
Moreover, both utilise
computer algebra packages to ensure minimal work for a user\footnote{Earlier 
publicly available numerical packages for the power spectrum in canonical multi-field inflation 
are \texttt{Pyflation} (\href{http://pyflation.ianhuston.net}{pyflation.ianhuston.net}  \cite{Huston:2011vt, Huston:2011fr}) 
and \texttt{MultiModeCode} \cite{Price:2014xpa}, and a publicly available code for the bispectrum in single 
field inflation 
is \texttt{BINGO} \cite{Hazra:2012yn}. Other numerical work at the level of the bispectrum in the single field case includes Refs.~\cite{Chen:2006xjb,Chen:2008wn,Horner:2013sea,Horner:2015gza}.}. The first package was developed by Seery, \texttt{CppTransport}, and represents a sophisticated set of tools developed in \texttt{C++} 
utilising the power of a number of external \texttt{C++} libraries, including 
\texttt{GiNaC} for front end computer algebra manipulations, and \texttt{BOOST} for the evolution of ODEs. It also contains a 
bespoke and sophisticated preprocessor, and automated data achieving and retrieval  tools. On the other hand, the second 
 package 
developed by Mulryne, \texttt{PyTransport}, is intended to be a more light weight product, 
built on a rather direct implementation of the transport framework. A working \texttt{Python} 
installation (with particular packages installed)
and a \texttt{C++} compiler are its only dependences. The core of \texttt{PyTransport} is written in \texttt{C++} to ensure 
good numerical performance, 
but the algebraic manipulations are handled by  \texttt{Python}'s \texttt{SymPy} package. 
Once an  inflationary model is specified, front end functions automatically edit \texttt{C++} code that is then 
complied into a bespoke \texttt{Python} module. This approach combines the speed of \texttt{C++} with the 
convenience of \texttt{Python}. Data storage and analysis are left to the user. By embedding the code in \texttt{Python}, however, the 
power of its many packages written for these purposes can be readily harnessed.  

A second aim of the present work, therefore, is to 
introduce a new version of the \texttt{PyTransport} package \texttt{PyTransport\,2.0}, which 
extends the code to the case of a non-trivial field-space metric.
Our new package allows users to specify both the potential and the field-space 
metric for a given model in a \texttt{Python} script, and automatically takes both these functions and 
generates a bespoke 
\texttt{Python} module. This module contains a number of useful functions 
including those needed to calculate the power spectrum and bispectrum of $\zeta$. The package is 
available at \href{https://github.com/jronayne/PyTransport}{github.com/jronayne/PyTransport}. 
Ref.~\cite{Mulryne:2016mzv} has also been updated such that version 2 details how to use this new code.  

Concurrently with our work, in an independent study 
Seery and Butchers have also extended the transport framework to the case of a non-Euclidean 
metric \cite{SeeryButchers}, and have incorporated their work into a new version of the \texttt{CppTransport} package, which 
is currently available as an experimental version at \href{https://github.com/ds283/CppTransport}{github.com/ds283/CppTransport}.

A non-trivial field-space metric is an important feature of inflationary models that 
arises in a number of contexts. First, it may be that 
a system of fields with a Euclidean metric may be more easily described in an 
alternative coordinate system. In this case 
the metric remains flat, but is nevertheless of a different form.
The second possibility is that the field-space metric is curved, which 
arises in many circumstances. Classic examples include 
when non-minimally coupled fields 
are rewritten as minimally coupled fields in the Einstein frame (see Refs.~\cite{Kaiser:2010ps,Greenwood:2012aj,Kaiser:2013sna}), and when
inflationary models are derived in supergravity. 
We note that a non-trivial field-space metric can be just as important as the fields' potential energy 
in determining the fields' 
dynamics, and hence the observational predictions of inflationary models. 

Our work is structured as follows. In the first part of the paper we follow the 
general framework set out in Ref.~\cite{Dias:2016rjq} closely and provide the  
additional calculations needed for our more general setting. 
First, in \S\ref{section2} we derive the second and third order action for the
covariant ``field'' perturbations first introduced in Ref.~\cite{Gong:2011uw} and 
subsequently used in Ref.~\cite{Elliston:2012ab}
to analytically study the bispectrum with a curved field-space metric (see also Ref.~\cite{Kaiser:2012ak}). 
Treating these perturbations and their canonical 
momenta as operators, we calculate Hamilton's equations of motion. Then we 
briefly review how Hamilton's equations can be used to calculate equations of motion for  
the correlations of the fluctuations in \S\ref{section3}. These equations are the transport 
equations which give our approach its 
name, and we provide them explicitly for the non-trivial field-space metric case. Finally, we calculate 
initial conditions for this system using the In-In formalism in \S\ref{section4}, and 
derive the relation between the covariant field perturbations and 
the curvature perturbation $\zeta$, which allows field-space correlations to be converted into correlations of $\zeta$, in 
\S\ref{section5}.  
This completes the specific equations needed to implement the framework of Ref.~\cite{Dias:2016rjq} 
for the case of a non-Euclidean field-space metric. We next turn to our numerical implementation of the equations we have derived 
in the \texttt{PyTransport\,2.0} package. After discussing briefly our implementation we showcase its 
utility with a number of examples in \S\ref{numerics}. We conclude in \S\ref{conclusions}.

\section{Perturbed action and Hamilton's equations}
\label{section2}
We begin by deriving the action to cubic order, and the 
Hamiltonian equations of motion, for covariant field-space perturbations defined on flat hypersurfaces. 
As we have discussed, the 
calculations mirror those presented in Ref.~\cite{Dias:2016rjq} but generalised to 
the case of a trivial field-space metric.

We begin with the action for ${\mathcal{N}}$ scalar fields minimally coupled to gravity 
\begin{equation}
S=\frac{1}{2}\int d^4x\sqrt{-g}\left[M_\mathrm{p}^2R-G_{IJ}g^{\mu\nu}\partial_{\mu}\phi^I\partial_{\nu}\phi^J-2V\right]\,,
\label{act1}
\end{equation}
where $R$ is the Ricci scalar associated 
with the spacetime metric $g_{\mu \nu}$, 
$G_{IJ}$ is the ${\mathcal{N}}$ dimensional 
field-space metric, and where upper case Roman indices run from $1$ to ${\mathcal{N}}$, which are 
raised and lowered by $G_{IJ}$.
$G_{IJ}$ is a function of the fields. 

For a flat Friedmann-Robertson-Walker (FRW) cosmology this action leads to the background equations of motion,
\begin{equation}
\begin{split}
3M_\mathrm{p}^2H^2=&\frac{1}{2}G_{IJ}\dot{\phi}^I\dot{\phi}^J+V\,,\\
D_t\dot{\phi}_I+3H\dot \phi_I=&-V_{I}\,,
\end{split}
\label{eq:background}
\end{equation}
where the covariant time derivative of a field-space vector, $U^I$, is defined as 
\begin{equation}
\label{covariantderT}
D_t U^I = \dot{U}^I+\dot\phi^M\Gamma^I_{MN}U^N\,,
\end{equation}
and $t$ indicates cosmic time, with a over-dot indicating differentiation with respect to cosmic time. 
The connection $\Gamma^I_{MN}$ is the Levi-Civita connection compatible with the field-space metric $G_{IJ}$.

We now consider perturbations about 
the FRW background. It proves convenient to follow 
Refs.~\cite{Maldacena:2002vr,Seery:2005wm,Seery:2005gb,Elliston:2012ab} and employ the (3+1) ADM 
decomposition of spacetime,
such that 
\begin{equation}
g_{00}=-(N^2-N_iN^i), \quad g_{0i}=N_i, \quad g_{ij}=h_{ij}\,,
\end{equation} 
where $N$ is the lapse function, $N_i$ is the shift vector, $h_{ij}$ the spatial metric, and lower case Roman indices run over the 
spatial coordinates.
With this choice of variables, the action (\ref{act1}) is written as
\begin{equation}
S=\frac{1}{2}\int d^4x\sqrt{h}\left(M_\mathrm{p}^2\left[NR^{(3)}+\frac{1}{N}(E_{ij}E^{ij}-E^2)\right]+\frac{1}{N}\pi^I\pi_I-NG_{IJ}\partial_i\phi^I\partial^i\phi^J-2NV\right),
\label{act2}
\end{equation}
where $R^{(3)}$ is the Ricci scalar of the 3-metric $h_{ij}$. The quantity $E_{ij}$ is proportional to the extrinsic curvature on 
slices of constant $t$, with 
\begin{equation}
E_{ij}=\frac{1}{2}(\dot{h}_{ij}-N_{i|j}-N_{j|i}),
\end{equation} 
where a bar denotes covariant derivatives with respect to the three metric. The quantity $\pi^I$ is defined as
\begin{equation}
\pi^I=\dot{\phi}^I-N^j\phi^I_{|j}.
\end{equation}

\subsection{Metric perturbations}

Working in the spatially flat gauge, and considering only scalar perturbations\footnote{Although beyond linear order vector and tensor perturbations do couple to the scalar perturbations, they do not affect the calculation of the scalar three point function which follows  from 
the third-order action involving only scalar perturbations.}, one has
$R^{(3)}=0$ and $h_{ij}=a^2\delta_{ij}$, and the only perturbations to the spacetime
metric are given by
\begin{equation}
\begin{split}
N&=1+\Phi_1+\Phi_2+\cdots \\
N_i&=\theta_{1\,,i}+\theta_{2\,,i}+\cdots\,,
\end{split}
\end{equation}
where $\Phi_1$ and $\Phi_2$ are the first and 
second order perturbations in the lapse, and $\theta_1$ and $\theta_2$ are the 
first and second order perturbations in the shift. 

\subsection{Field perturbations}
Next we consider the perturbations to the matter sector and hence to the 
scalar fields present.  The field perturbations,  $\delta \phi^I(x,t)$,  are defined  by
the expression  
$\phi^I = \phi_0^I(t) + \delta \phi^I(x,t)$. These field-space perturbations are 
not, however, covariant under relabelling of field-space, and 
it proves convenient  to work with 
a different set of perturbations that are covariant, which we label $Q^I$. These were first 
introduced by Gong \& Tanaka \cite{Gong:2011uw}. The idea is to consider the 
geodesic that links together the position in field-space labelled by $\phi^I_0$ and that labelled  
by $\phi^I$, and an affine parameter parametrising this trajectory denoted $\lambda$. The 
coordinate displacement $\delta \phi^I$ can then be expressed by the series expansion about the point $\lambda=0$ as
\begin{equation}
\delta\phi^I= \left.{\frac{d\phi^I}{d\lambda}}\right|_{\lambda=0}+\left.\frac{1}{2!}\frac{d^2\phi^I}{d\lambda^2}\right|_{\lambda=0}+\cdots \,.
\label{taylor}
\end{equation} 
We can then form the geodesic equation
\begin{equation}
D_{\lambda}^2\phi^I=\frac{d^2\phi^I}{d\lambda^2}+\Gamma^{I}_{JK}\frac{d\phi^J}{d\lambda}\frac{d\phi^K}{d\lambda}=0\,,
\label{geo}
\end{equation}
and define $Q^I={\rm d} \phi^I/{\rm d}\lambda|_{\lambda =0}$ and 
$D_{\lambda}=Q^I\nabla_I$ (where $\nabla_I$ is the covariant derivative).
Using this geodesic equation, the expansion (\ref{taylor}) 
can be rewritten as 
\begin{equation}
\delta\phi^I={Q^I}-\frac{1}{2!}\Gamma^I_{JK}Q^JQ^K\,,
\label{QdPhi}
\end{equation}
which relates field perturbations to the covariant perturbations.
The time derivative of field fluctuations, $\delta \dot \phi^I$, can also be written in terms covariant 
quantities as 
\begin{equation}
\delta \dot\phi^I 
 =D_t Q^I -\dot{\phi}^M\Gamma^I_{MN}Q^N -\frac{1}{2} \Gamma^I_{JK,M}\dot{\phi}^MQ^JQ^K -\Gamma^I_{(JK)}D_tQ^JQ^K  + \Gamma^I_{(JK)}\Gamma^J_{MN}Q^K\dot{\phi}^MQ^N\,,
\label{dQdPhi}
\end{equation}
as can a perturbation to the field-space metric, and using (\ref{QdPhi}) we find
\begin{eqnarray}
\delta G_{IJ}
&=& 2\Gamma_{(IJ)K} Q^K -\Gamma_{(IJ)K}\Gamma^K_{MN}Q^MQ^N + \Gamma_{(IM)L}\Gamma^M_{JK}Q^K Q^L + \Gamma_{(JM)L}\Gamma^M_{IK}Q^K Q^L\nonumber \\
 &~& +\, \frac{1}{2}(G_{IM}\Gamma^M_{JK,L} + G_{JM}\Gamma^M_{IK,L})Q^K Q^L\,.\label{dGdPhi}
\end{eqnarray}
Here we have adopted the notation of using $(IJ)$ parenthesis to illustrate symmetrization over the indices $I$ and $J$. A bar $|$ is used to excluded certain indices from the symmetrization procedure, for example, $(I|J|K)$ symmetrizes $I$ and $K$ but not $J$.

\subsection{The perturbed action}
\label{act}
The next step is to insert our perturbed expressions for $N$, $N_i$ and $\phi^I$ 
into  (\ref{act2}) to calculate the perturbed action.
Expanding order by order, the first order action simply leads back to the background equations, while 
the action at second and higher order lead to the dynamics of the perturbations. 

After some integration by parts and discarding total  derivatives, one finds 
the action at second and third order 
can be written in the form given by Elliston {\it et al.} \cite{Elliston:2012ab}
\begin{multline}
\label{Act2}
S_{(2)}=\frac{1}{2}\int d^4 x a^3 \left( \Phi_1\left[-6M_\mathrm{p}^2 H^2\Phi_1 +G_{IJ}\dot{\phi}^I\dot{\phi}^J\Phi_1 \right. \right. \\ \left. 
-2G_{IJ}\dot{\phi}^ID_tQ^J-2V_{;I}Q^I\right]- \frac{2}{a^2}\partial^2\theta_1\left[2M_\mathrm{p}^2H\Phi_1 -G_{IJ}\dot{\phi}^IQ^J 
\right] \\ \left.
+ R_{KIJL}\dot{\phi}^K\dot{\phi}^LQ^IQ^J + G_{IJ}D_tQ^ID_tQ^J -G_{IJ}\partial^iQ^I\partial_jQ^J - V_{;IJ}Q^IQ^J\right)\,,
\end{multline}
and
\begin{multline}
\label{Act3}
S_{(3)}=\frac{1}{2}\int d^4x a^3\left( 6M_\mathrm{p}^2H^2\Phi_1^3 + 4M_\mathrm{p}^2\frac{H}{a^2}\Phi_1^2\partial^2\theta_1-\frac{M_\mathrm{p}^2\Phi_1}{a^4}(\partial_i\partial_j\theta_1\partial_i\partial_j\theta_1 - \partial^2\theta_1\partial^2\theta_1) \right.
\\ \left. -G_{IJ}\dot{\phi}^I\dot{\phi}^J\Phi_1^3+2\Phi_1^2\dot{\phi}^ID_tQ^J 
+ \frac{2}{a^2}\Phi_1G_{IJ}\dot{\phi}^I\partial_i\theta_1\partial_iQ^J - \Phi_1R_{L(IJ)M}\dot{\phi}^L\dot{\phi}^MQ^IQ^J \right. \\ \left. - \Phi_1\left(G_{IJ}Q^IQ^J +\frac{1}{a^2}G_{IJ}\partial^iQ^I\partial_jQ^J \right) -\frac{2}{a^2}\partial_i\theta_1G_{IJ}D_tQ^I\partial_iQ^J 
+ \frac{4}{3}R_{I(JK)L}\dot{\phi}^LD_tQ^IQ^JQ^K \right. \\ \left.
+ \frac{1}{3}R_{(I|LM|J;K)}\dot{\phi}^L\dot{\phi}^MQ^IQ^JQ^K - \frac{1}{3}V_{;(IJK)}Q^IQ^JQ^K-V_{;(IJ)}\Phi_1Q^IQ^J \right),
\end{multline}
where $R_{IJKL}$ is the Riemann tensor compatible with the field-space metric $G_{IJ}$, and $R_{IJKL;M}$ 
it's covariant derivative. 
\subsubsection{Constraint equations}

Varying the action with respect to the lapse and shift leads to two 
constraint equations that can be used to provide 
expressions for the perturbations in the lapse and shift in terms of the covariant $Q^I$ 
perturbations
\cite{Langlois:1994ec}. 
These can be substituted back into the action to express the perturbed action only in terms of 
$Q^I$. To do so we only need the
constraint equations at linear order (as explained in \cite{Maldacena:2002vr}), but later we 
will also need them 
at second order too, so we provide the full expressions here. 

Considering first variation with respect to the shift, at linear order one finds 
\begin{equation}
\label{P1}
\Phi_1=\frac{1}{2M_\mathrm{p}^2H}G_{IJ}\dot{\phi}^IQ^J,
\end{equation}
while at second order
\begin{equation}
\label{P2}
\begin{split}
\Phi_2 = \frac{\Phi_1^2}{2}+\frac{\partial^{-2}}{2M_\mathrm{p}^2H}&\left[  -\frac{M_\mathrm{p}^2}{a^2}\partial_i\partial_j\Phi_1\partial_i\partial_j\theta_1 + \frac{M_\mathrm{p}^2}{a^2}\partial^2 \Phi_1 \partial^2\theta_1 \right. \\
&\left. + G_{IJ}(\partial_iD_tQ^I)\partial_iQ^J + G_{IJ} D_t Q^I \partial^2 Q^J\right].
\end{split}
\end{equation}
On large scales where spatial gradients decay, one then finds that
\begin{equation}
\Phi_2 = \frac{\Phi_1^2}{2}+\frac{\partial^{-2}}{2M_\mathrm{p}^2H}\left[G_{IJ}(\partial_iD_tQ^I)\partial_iQ^J + G_{IJ} D_t Q^I \partial^2 Q^J\right]\,.
\end{equation}
Next varying the action with respect to the lapse, at linear order we have 
\begin{equation}
\label{T1}
\partial^2\theta_1 = -3a^2H\Phi_1+\frac{a^2}{2M_\mathrm{p}^2H}G_{IJ}\Phi_1\dot{\phi}^I\dot{\phi}^J-\frac{a^2}{2M_\mathrm{p}^2H}G_{IJ}\dot{\phi}^ID_tQ^J-\frac{a^2}{2M_\mathrm{p}^2 H}V_{;I}Q^I\,,
\end{equation}
and at second order
\begin{equation}
\begin{split}
\partial^2\theta_2 = & 2\Phi_1\partial^2\theta_1 - \frac{1}{4a^2H}\left(\partial_i\partial_j\theta_1\partial_i\partial_j\theta_1-\partial^2\theta_1\partial^2\theta_1 \right) + \frac{a^2}{2M_\mathrm{p}^2 H}G_{IJ}\Phi_1\dot{\phi}^ID_tQ^J \\ &
+ \frac{1}{2M_\mathrm{p}^2H}G_{IJ}\dot{\phi}^I\partial_i\theta_1\partial_i Q^J - \frac{a^2}{4M_\mathrm{p}^2}G_{IJ}D_tQ^ID_tQ^J - \frac{1}{4M_\mathrm{p}^2H} G_{IJ}\partial_i Q^I \partial_i Q^J 
\\ &
-  \frac{a^2}{4M_\mathrm{p}^2}V_{;(IJ)}Q^IQ^J + \frac{a^2 H}{2}(2\Phi_2 - 3 \Phi_1^2)(\epsilon -3) -  \frac{a^2}{4M_\mathrm{p}^2} R_{L(IJ)M}\dot{\phi}^L\dot{\phi}^MQ^IQ^J \,,
\end{split}
\end{equation}
where  $\epsilon = -\dot{H}/H^2$ is the slow-roll parameter. 
Using these latter expressions and again taking the large scale superhorizon limit one finds the additional 
relation
\begin{equation}
\label{ham1}
6H\Phi_1=\frac{1}{M_\mathrm{p}^2H}G_{IJ}\Phi_1\dot{\phi}^I\dot{\phi}^J-\frac{1}{M_\mathrm{p}^2H}G_{IJ}\dot{\phi}^ID_tQ^J-\frac{1}{M_\mathrm{p}^2 H}V_{;I}Q^I \,,
\end{equation}
at first order, and 
\begin{equation}
\begin{split}
\label{constraint2}
\frac{1}{2}G_{MN}D_tQ^MD_tQ^N =& 2 \Phi_1 G_{IN}\dot{\phi}^ID_t Q^N - \frac{1}{2}V_{;(MN)}Q^MQ^N \\&-M_\mathrm{p}^2 H^2 ( 3\Phi_1^2 - 2\Phi_2)(\epsilon - 3) - \frac{1}{2} R_{I(MN)J}\dot{\phi}^I\dot{\phi}^JQ^MQ^N\,,
\end{split}
\end{equation}
at second order. 

\subsubsection{The Fourier space action}

Finally, using the equations for  $\Phi$ (\ref{P1}) and $\theta$ (\ref{T1}) 
in terms of $Q^I$ one can write the 
quadratic and cubic parts of the action (\ref{Act2}) and (\ref{Act3}) solely in 
terms of $Q^I$. 
It is convenient at this stage to move from real space to Fourier space. 
After doing so, to 
keep our expressions to a manageable size, 
we follow the extended summation convention introduced in 
Ref.~\cite{Dias:2016rjq}. When considering 
Fourier space quantities we use 
bold font indices, $\bf{I},\bf{J},\dots$ to indicate that the 
usual summation over fields is accompanied by an integration over Fourier space. 
For example, 
\be
A^{\bf I}B_{\bf I}  = \int \frac{ d^3k_I }{(2\pi)^3} A^{I}({\bf k}_I) B_I({\bf k}_I)\,,
\ee
where the subscript $I$ on ${\bf k}_I$ indicates that this is the 
wavenumber associated with objects that carry the $I$ index. 
Using this notation the action reads
\begin{equation}
\label{S2k}
S_{(2)}= \frac{1}{2}\int dt   a^3\left( G_{\bf{IJ}}({\bf{k}}_I,{\bf{k}}_J)( D_tQ^{\bf{I}}({\bf{k}}_I)D_tQ^{\bf{J}}({\bf{k}}_J) + M_{\bf{IJ}}({\bf{k}}_I,{\bf{k}}_J)Q^{\bf{I}}({\bf{k}}_I)Q^{\bf{J}}({\bf{k}}_J) \right) \,,
\end{equation}
at second order and
\begin{equation}
\begin{split}
\label{S3k}
S_{(3)}=\frac{1}{2}\int dt a^3
&\left( A_{\bf{IJK}}({{{\bf{k}}_I,{\bf{k}}_J,{\bf{k}}_K}})Q^{\bf{I}}({\bf{k}}_I)Q^{\bf{J}}({\bf{k}}_J)Q^{\bf{K}}({\bf{k}}_K)\right. \\ &\left.
+ B_{\bf{IJK}}({{{\bf{k}}_I,{\bf{k}}_J,{\bf{k}}_K}})D_tQ^{\bf{I}}({\bf{k}}_I)Q^{\bf{J}}({\bf{k}}_J)Q^{\bf{K}}({\bf{k}}_K)\right. \\ &\left.
+ C_{\bf{IJK}}({{{\bf{k}}_I,{\bf{k}}_J,{\bf{k}}_K}})D_tQ^{\bf{I}}({\bf{k}}_I)D_tQ^{\bf{J}}({\bf{k}}_J)Q^{\bf{K}}({\bf{k}}_K) \right)\,,
\end{split}
\end{equation}
at third order, where we have defined 
\begin{eqnarray}
G_{\bf{IJ}}({{\bf{k}}_I,{\bf{k}}_J})& &=(2\pi)^3\delta{({\bf{k}}_I+{\bf{k}}_J)} G_{IJ}\\
M_{\bf{IJ}}({{\bf{k}}_I,{\bf{k}}_J})& &=(2\pi)^3\delta{({\bf{k}}_I+{\bf{k}}_J)} \left(\frac{k_I^2}{a^2}G_{IJ}-m_{IJ} \right)
\label{eq:M}
\\
A_{\bf{IJK}}({{\bf{k}}_I,{\bf{k}}_J,{\bf{k}}_K})& &=(2\pi)^3\delta{({\bf{k}}_I+{\bf{k}}_J+{\bf{k}}_K)}a_{IJK}\\
B_{\bf{IJK}}({{\bf{k}}_I,{\bf{k}}_J,{\bf{k}}_K})& &=(2\pi)^3\delta{({\bf{k}}_I+{\bf{k}}_J+{\bf{k}}_K)}b_{IJK}\\
C_{\bf{IJK}}({{\bf{k}}_I,{\bf{k}}_J,{\bf{k}}_K})& &=(2\pi)^3\delta{({\bf{k}}_I+{\bf{k}}_J+{\bf{k}}_K)}c_{IJK}\,.
\end{eqnarray}
with 
\begin{equation}
m_{IJ}= V_{;IJ} -R_{IKLJ}\dot{\phi}^K\dot{\phi}^L -\frac{3+\epsilon}{M_\mathrm{p}^2}\dot{\phi}_i\dot{\phi}_J - \frac{(\dot{\phi}_I D_t\dot{\phi}_J +\dot{\phi}_J D_t\dot{\phi}_I)}{HM_\mathrm{p}^2}\,,
\end{equation}
and
\begin{equation}
\begin{split}
\label{aijk}
a_{IJK}=& -\frac{1}{3}V_{;IJK}-\frac{\dot{\phi}_IV_{;JK}}{2HM_\mathrm{p}^2}+\frac{\dot{\phi}_I\dot{\phi}_J\xi_K}{8H^2M_\mathrm{p}^4} + \frac{\dot{\phi}_I\xi_J\xi_K}{32H^3M_\mathrm{p}^4}\left(1 - \frac{({{\bf{k}}_J \cdot {\bf{k}}_K})^2}{k_J^2k_K^2} \right) \\ 
&+\frac{\dot{\phi}_I\dot{\phi}_J\dot{\phi}_K}{8HM_\mathrm{p}^4}\left(6 \frac{G_{MN}\dot{\phi}^M\dot{\phi}^N}{H^2M_\mathrm{p}^2}\right) +\frac{\dot{\phi}_IG_{JK}}{2HM_\mathrm{p}^2}\frac{{{\bf{k}}_J \cdot {\bf{k}}_K}}{a^2} \\
 &- \frac{1}{2}\frac{G_{NK}\dot{\phi}^L\dot{\phi}^M\dot{\phi}^NKR_{L(IJ)M}}{M_\mathrm{p}^2H}+\frac{1}{3}\dot{\phi}^L\dot{\phi}^MR_{(I|LM|J;K)}\,,
\end{split}
\end{equation}

\begin{equation}
\begin{split}
\label{bijk}
b_{IJK}=\frac{\dot{\phi}_I\dot{\phi}_J\dot{\phi}_K}{4H^2M_\mathrm{p}^4}-\frac{\dot{\phi}_I\xi_J\dot{\phi}_K}{8H^3M_\mathrm{p}^4}\left(1 - \frac{({{\bf{k}}_J \cdot {\bf{k}}_K})^2}{k_J^2k_K^2} \right) -\frac{\xi_IG_{JK}}{2HM_\mathrm{p}^2}\frac{{{\bf{k}}_I \cdot {\bf{k}}_J}}{k^2_I} + \frac{4}{3}\dot{\phi}^L R_{I(JK)L}\,, 
\end{split}
\end{equation}

\begin{equation}
\begin{split}
\label{cijk}
c_{IJK}=-\frac{G_{IJ}\dot{\phi}_K}{2HM_\mathrm{p}^2}+ \frac{\dot{\phi}_I\dot{\phi}_J\dot{\phi}_K}{8H^3M_\mathrm{p}^4}\left(1-\frac{({{\bf{k}}_I \cdot {\bf{k}}_J})^2}{k_I^2k_J^2}\right) + \frac{G_{IJ}\dot{\phi}_K}{HM_\mathrm{p}^2}\frac{{{\bf{k}}_I \cdot {\bf{k}}_K}}{k^2_I}\,, \\
\end{split}
\end{equation}
where 
\begin{equation}
\xi_I=2 D_t \dot{\phi}_I+\frac{\dot{\phi}_I}{H}\frac{G_{NM}\dot{\phi}^N\dot{\phi}^M}{M_\mathrm{p}^2}\,.
\end{equation}
Here $a_{IJK}$ is to be symmetrised over all three indices, $b_{IJK}$  over $J$ \& $K$ and $c_{IJK}$  over $I$ \& $J$. Each index permutation will have a corresponding 
exchange of wavenumber associated with the indices.

\subsection{Hamilton's equations}
\label{hamfor}

From the action we can derive equations of motion for the perturbations $Q^I(\mathbf{k})$. Perturbations 
behave quantum mechanically on subhorizon scales, and to account 
for this we introduce the conjugate momenta to $Q^I$, $P^I$, and treat $Q^I$ and $P^I$ as 
Heisenberg picture operators which obey Hamilton's equations.

The canonical momentum is defined as
\begin{equation}
P_I=\frac{\delta S}{\delta(D_t Q^I)}\,,
\end{equation}
and obeys the relation,
\begin{equation}
\left[Q^I({\bf{k}}_I,t),P_J({\bf{k}}_J,t')\right]=i(2\pi)^3\delta^I_J({\bf{k}}_I+{\bf{k}}_J)\delta(t-t')\,.
\end{equation}
Utilising Eqs.~(\ref{S2k}) \& (\ref{S3k}) one finds
\begin{equation}
P_I=a^3\left(D_tQ_I + \frac{1}{2}B_{\bf JKI}Q^{\bf J}Q^{\bf K}+C_{\bf IJK}P^{\bf J}Q^{\bf K}\right)\,.
\end{equation}

At this stage it is helpful to rescale $P_I$ such that $P_I\rightarrow a^3P_I$, 
where for convenience we employ the same symbol for the rescaled momentum, and use it solely from 
here on. In terms of the rescaled momentum 
\begin{equation}
D_tQ_I=P_I-\frac{1}{2}B_{\bf JKI}Q^{\bf J}Q^{\bf K}-C_{\bf IJK}P^{\bf J}Q^{\bf K}+\cdots.
\end{equation}
The Hamiltonian is then given by
\begin{equation}
\begin{split}
\mathcal{H} &=  \int dt\frac{a^3}{2}\left(\underbrace{G_{\bf IJ}P^{\bf I}P^{\bf J} - M_{\bf IJ}Q^{\bf I}Q^{\bf J}}_{\mathcal{H}_0}\right.\\ & \hspace{5cm}\left.-\underbrace{A_{\bf IJK}Q^{\bf I}Q^{\bf J}Q^{\bf K}-B_{\bf IJK}Q^{\bf I}Q^{\bf J}P^{\bf K} - C_{\bf IJK}P^{\bf I}P^{\bf J}Q^{\bf K}}_{\mathcal{H}_{int}}\right)\,,
\label{hamtot}
\end{split}
\end{equation}
where have labelled the `free' part of the Hamiltonian
$\mathcal{H}_0$, and the interaction part, $\mathcal{H}_{\rm int}$.

Finally Hamilton's equations provide us with evolution equations for $Q^I$ and $P^I$ 
which are
\begin{equation}
\label{Ham1}
D_tQ^I=-i[Q^I,\mathcal{H}]
\end{equation}
\begin{equation}
\label{Ham2}
D_tP^I=-i[P^I,\mathcal{H}]- 3HP^I\,,
\end{equation}
where the evolution of $P^I$ takes a slightly non-canonical form due to the rescaling of the 
canonical momenta. 

\section{The transport equations}
\label{section3}
Once equations of motion are known for the Heisenberg operators, these 
can immediately be converted into equations of motion for expectation values of 
products of these operators using 
Ehrenfest's theorem \cite{Mulryne:2013uka} . This is the idea behind the Transport approach and 
was explored in detail in Ref.~\cite{Dias:2016rjq}, where the reader can turn for further details.
For convenience, we first label the full 
phase space of Heisenberg operations with the symbol $\delta X^a$, where 
$\delta X^a = (Q^I, P^J)$
and where lower case Roman indices run from 
$1$ to $2{\cal N}$.  The expectation values we are 
interested in are then the two and three-point functions 
of $\delta X^a$ 
\begin{eqnarray}
  \langle \delta X^{a}({\bf k}_a) \delta X^{b}({\bf k}_b) \rangle
  &=&
  (2\pi)^3 \delta({\bf{k}}_a + {\bf{k}}_b) \Sigma^{ab}(k_a) 
  \label{eq:sigma-def} \\
    \langle \delta X^{{a}}({\bf k}_a) \delta X^{{b}}({\bf k}_b) \delta X^{{c}}({\bf k}_c) \rangle
 &=&
  (2\pi)^3 \delta({\bf{k}}_a + {\bf{k}}_b + {\bf{k}}_c) B^{abc}(k_a, k_b, k_c) .
  \label{eq:B-def}
\end{eqnarray}
As described, the equations of motion for these correlation functions follow directly from Eqs.~(\ref{Ham1})-(\ref{Ham2})
together with Ehrenfest's theorem, and can be presented in terms of equations of motion 
for $\Sigma^{ab}$ and $B^{abc}$. In our covariant setting these take the form 
\begin{equation}
 D_t \Sigma^{ab}(k)
  =
  {u}^a{}_c(k) \Sigma^{cb}(k) + {u}^b{}_c(k) \Sigma^{ac}(k) ,
  \label{eq:twopf-transport-ode}
  \end{equation}
  and
  \begin{equation}
\begin{split}
  D_t B^{abc}(k_a, k_b, k_c)
  = \mbox{} &
  {u}^a{}_d(k_a) B^{dbc}(k_a, k_b, k_c)
  +
  {u}^b{}_d(k_b) B^{adc}(k_a, k_b, k_c)
  +
  {u}^c{}_d(k_c) B^{abd}(k_a, k_b, k_c)
  \\
  & \mbox{}
  +
  {u^a}_{de}({\bf{k}}_a, -{\bf{k}}_b, -{\bf{k}}_c)
  \Sigma^{db}(k_b) \Sigma^{ec}(k_c)
  \\
  & \mbox{}
  +
  {u^b}_{de}({\bf{k}}_b, -{\bf{k}}_a, -{\bf{k}}_c)
  \Sigma^{ad}(k_a) \Sigma^{ec}(k_c)
  \\
  & \mbox{}
  +
  {u^c}_{de}({\bf{k}}_c, -{\bf{k}}_a, -{\bf{k}}_b)
  \Sigma^{ad}(k_a) \Sigma^{be}(k_c) \,,
  \label{eq:threepf-transport-raw-ode}
\end{split}
\end{equation}
where the covariant time derivative acts on $\Sigma^{ab}$ in the following way
\begin{equation}
D_t \Sigma^{ab}(k) = \partial_t\Sigma^{ab}(k) +  {\bf{\Gamma}}^a_c(k)\Sigma^{cb}(k) +  {\bf{\Gamma}}^b_c(k)\Sigma^{ac}(k)\,,
\end{equation}
and on $B^{abc}$ as 
\begin{equation}
\begin{split}
D_tB^{abc}(k_a, k_b, k_c) =& \partial_t B^{abc}(k_a, k_b, k_c) +  {\bf{\Gamma}}^a_d(k)B^{dbc}(k_a, k_b, k_c)\\
&+  {\bf{\Gamma}}^b_d(k)B^{adc}(k_a, k_b, k_c)+  {\bf{\Gamma}}^c_d(k)B^{abd}(k_a, k_b, k_c)\,,
\end{split}
\end{equation}
with  ${\bf{\Gamma}}^a_b$ is defined as
\begin{equation}
{\bf{\Gamma}}^a_b	=\left(
		\begin{array}{cc}
			\Gamma^I_{JK}\dot{\phi}^K & 0 \\
			0 & \Gamma^I_{JK}\dot{\phi}^K 
		\end{array}
	\right) \,,
\end{equation}
The $u$-tensors take the 
form
\begin{equation}
	{u^{a}}_{b}
	=
	\left(
		\begin{array}{cc}
			0 & \delta^{{I}}_{J} \\
			{\tilde m^{{I}}}_{J} & - 3H \delta^{{I}}_{J}
		\end{array}
	\right) \,,
		\label{eq:u2-general}
\end{equation}
where 
\begin{equation}
\tilde m_{IJ} = -\frac{k^2}{a^2} G_{IJ} - m_{IJ}\,,
\end{equation}
and
\begin{equation}
	{u^a}_{bc}
	=
	\left\{
		\begin{array}{c}
			\left(
			\begin{array}{c@{\hspace{3mm}}c}
				- {b_{JK}}^{{I}} & -{c^{{I}}}_{JK} \\
				3 {a^{{I}}}_{JK} & {b^{{I}}}_{KJ}
			\end{array}
			\right)
			\\
			\\
			\left(
			\begin{array}{c@{\hspace{3mm}}c}
				- {c^{{I}}}_{KJ} & 0 \\
				{b^{{I}}}_{JK} & {c_{KJ}}^{{I}}
			\end{array}
			\right)
		\end{array}
	\right\}\,.
	\label{eq:u3-general}
\end{equation}
\,
\subsection{Transport equations for real valued quantities}
 
The two-point function will in general be complex, and can be 
divided into its real and imaginary parts 
\be
\Sigma^{ad}=\Sigma^{ad}_{\text{\tiny Re}}+i\Sigma^{ad}_{\text{\tiny Im}}\,,
\ee
with the real part symmetric under interchange of its indices, and the 
imaginary part anti-symmetric. Both parts independently satisfy Eq.~(\ref{eq:twopf-transport-ode}). 
On superhorizon scales the imaginary part 
decays to zero, indicating that 
on large scales the statistics of inflationary perturbations follow classical equations of motion. 

$B^{abc}$, is in general also complex, but is 
real when only tree-level effects are included. 
In our numerical implementation of the transport system we evolve  
the real and imaginary parts of $\Sigma^{ab}$ separately using  Eq.~(\ref{eq:twopf-transport-ode}), 
and evolve $B^{abc}$ according to the equation
\begin{equation}
\begin{split}
  D_t B^{abc}(k_a, k_b, k_c)  =  &
  {u}^a{}_d(k_a) B^{dbc}(k_a, k_b, k_c)
  +
  {u}^b{}_d(k_b) B^{adc}(k_a, k_b, k_c)
  +
  {u}^c{}_d(k_c) B^{abd}(k_a, k_b, k_c)
  \\
  & \mbox{}
  +
  {u}^a{}_{de}(\vect{k}_a, \vect{k}_b, \vect{k}_c)
  \Sigma_{\text{\tiny Re}}^{db}(k_b) \Sigma_{\text{\tiny Re}}^{ec}(k_c)
  -
  {u}^a{}_{de}(\vect{k}_a, \vect{k}_b, \vect{k}_c)
  \Sigma_{\text{\tiny Im}}^{db}(k_b) \Sigma_{\text{\tiny Im}}^{ec}(k_c)
  \\
  & \mbox{}
  +
  {u}^b{}_{de}(\vect{k}_b, \vect{k}_a, \vect{k}_c)
  \Sigma_{\text{\tiny Re}}^{ad}(k_a) \Sigma_{\text{\tiny Re}}^{ec}(k_c)
  -
  {u}^b{}_{de}(\vect{k}_b, \vect{k}_a, \vect{k}_c)
  \Sigma_{\text{\tiny Im}}^{ad}(k_a) \Sigma_{\text{\tiny Im}}^{ec}(k_c)
  \\
  & \mbox{}
  +
  {u}^c{}_{de}(\vect{k}_c, \vect{k}_a, \vect{k}_b)
  \Sigma_{\text{\tiny Re}}^{ad}(k_a) \Sigma_{\text{\tiny Re}}^{be}(k_b) 
  -
  {u}^c{}_{de}(\vect{k}_c, \vect{k}_a, \vect{k}_b)
  \Sigma_{\text{\tiny Im}}^{ad}(k_a) \Sigma_{\text{\tiny Im}}^{be}(k_b) ,
\end{split}
\label{eq:threepf-transport-real-ode}
\end{equation}
which follows from Eq.~\ref{eq:threepf-transport-raw-ode} once $\Sigma^{ab}$ is broken into real and imaginary parts, and 
which makes it clear that $B^{abc}$ remains real if its initial conditions are real.

\section{Initial conditions for the two and three-point functions}
\label{section4}

In order to solve for $\Sigma^{ab}_{\text{\tiny Re}}$, $\Sigma^{ab}_{\text{\tiny Im}}$ and $B^{abc}$ 
numerically, the last element we need are initial 
conditions. 
Following the approach of Ref.~\cite{Dias:2016rjq} (which is closely related to  
that of Ref.~\cite{Chen:2008wn}), these are fixed at some 
early time at which all the wavenumbers of a given correlation are far inside the horizon during inflation, and 
where $m_{IJ}$ is subdominant to  $(k/a)^2 G_{IJ}$ in Eq.~(\ref{eq:M}). In this limit it is reasonable to assume that 
the solution for the two-point correlation function of $Q^I$ is well approximated by the de-Sitter space solution  
and we can use this solution to provide initial conditions for our numerical evolution. We note that it is only 
required that this solution be valid at some point long before all scales of interest cross the horizon, and 
moreover,  that the numerical evolution is then free to evolve away from this solution, accounting 
for the complex dynamics that can subsequently occur in general inflationary models. 

The two-point function in de Sitter space is typically written in conformal time $\tau$ and takes the form,
\begin{equation}
\langle Q^I (k_1,\tau_1)Q^J(k_2,\tau_2)\rangle = (2\pi)^3\delta(k_1+k_2)\Pi^{IJ} \frac{H^2}{{2 k^3}} (1-ik\tau_1)(1+ik\tau_2)e^{ik(\tau_1-\tau_2)}\,,
\label{IC2pt}
\end{equation}
where $\Pi^{IJ}$ is given by \cite{Elliston:2012ab}
\begin{equation}
\Pi^{IJ}(\tau_1,\tau_2)= {\mathcal{T}} \exp \left(-\int^{\tau_2}_{\tau_1}d\tau \Gamma^I_{KL}\left[\phi^M(\tau)\right]\frac{d\phi^K}{d\tau}\right)G^{LJ}(\tau_1)\,,
\end{equation}
which transforms as a bitensor with the first index $I$ transforming in the tangent space at point $\phi^M(\tau_2)$ and the second 
index $J$ in the tangent space at point $\phi^M(\tau_1)$. The two-point functions $\langle Q^I(\tau_1) P^J (\tau_2) \rangle$, and 
$\langle P^I (\tau_1) P^J(\tau_2)\rangle$ can then be calculated by differentiating Eq.~(\ref{IC2pt}), using the definition of $P^I$ 
and accounting for the 
use of conformal time. 
For our purposes we only need to consider 
the limit $\tau_2 \to \tau_1 $ with $-\tau \gg 1$, which corresponds to equal time correlations on sub-horizon scales.
In this limit $\Pi^{IJ} \to G^{IJ}$, and one finds
\begin{align}
\Sigma^{ab}_{*Re}&=	\frac{1}{2a^3k}\left(
		\begin{array}{cc}
			aG^{IJ} & -aHG^{IJ} \\
			-aHG^{IJ}& (k^2/a)G^{IJ}
		\end{array}
	\right)\\[1ex]
\Sigma^{ab}_{*Im}&=	\frac{1}{2a^3k}\left(
		\begin{array}{cc}
			0 & kG^{IJ} \\
			-kG^{IJ}& 0
		\end{array}
	\right)\,,
\end{align}
where we denote values at the initial time long before horizon crossing 
with an asterisk. The initial conditions for  $\Sigma^{ab}_{\text{\tiny Re}}$ where also 
given by Dias, Frazer and Seery \cite{Dias:2015rca}. Some further details are given in  appendix~\ref{apdxA}.

In order to calculate the initial conditions for $B^{abc}$ we need to calculate the three-point correlation functions  
for $Q^I$ and $P^I$ for Fourier modes on sub-horizon scales. As argued in Ref.~\cite{Dias:2016rjq}, these can be calculated using 
the In-In formalism. By writing the interaction part of the Hamiltonian given in Eq.~(\ref{hamtot}) in the form 
$\mathcal{H}_{int}=\mathcal{H}_{\bf abc}\delta X^{\bf a}\delta X^{\bf b}\delta X^{\bf c}$, the general expression 
for the three-point function can compactly be written as 
\begin{equation}
\langle \delta X^a \delta X^b \delta X^c\rangle_* = -i\int^{\tau_{init}}_{-\infty} {\rm d} \tau \left \langle \left[\delta X_*^a\delta X_*^b \delta X_*^c, \mathcal{H}_{\bf efg} \delta X^{\bf e} \delta X^{\bf f} \delta X^{\bf g}\right] \right \rangle\,,
\end{equation}
which leads to 
\begin{equation}
\label{intialB}
B^{abc}_* = -6i\int^{\tau_{init}}_{-\infty} d\tau \mathcal{H}_{efg}\Sigma^{ae}(\tau_*,\tau)\Sigma^{bf}(\tau_*,\tau)\Sigma^{cg}(\tau_*,\tau) +c.c. \,,
\end{equation}
where we have defined $\mathcal{H}_{abc}$ as
\begin{equation}
	\mathcal{H}_{abc}
	=\frac{1}{3!}
	\left\{
		\begin{array}{c}
			\left(
			\begin{array}{c@{\hspace{3mm}}c}
				- 3a_{IJK} & -b_{IKJ} \\
				 -b_{KJI} & -c_{IJK}
			\end{array}
			\right)
			\\
			\\
			\left(
			\begin{array}{c@{\hspace{3mm}}c}
				-b_{IJK} & c_{KJI} \\
				-c_{IKJ} & 0
			\end{array}
			\right)
		\end{array}
	\right\}\, ,
	\label{eq:hintabc}
\end{equation}
and $\Sigma^{ab}(\tau_1,\tau_2)$ with dependence on two times as
\begin{equation}
\label{twoTimes}
\langle \delta X^a (k_1,\tau_1)\delta X^b(k_2,\tau_2)\rangle = (2\pi)^3\delta({ \bf k}_1+{\bf k}_2)\Sigma^{ab}(\tau_1,\tau_2)\,.
\end{equation}
The explicit integrals which result for the different elements of $B^{abc}$ are similar in structure to 
those of the canonical field-space metric case presented in  Ref.~\cite{Dias:2016rjq}, where one 
can turn for a full discussion. When 
performing the integrations explicitly we must understand the time dependence of the terms 
which enter. 
The time dependence of the $b_{IJK}$ and $c_{IJK}$ tensors which appear 
in the  interaction Hamiltonian is slow-roll suppressed and their 
time dependence can be neglected. 
On the other hand, the $a_{IJK}$ tensor contains `fast' changing terms 
proportional to $(k/a)^2\sim (k\tau)^2$ which grow exponentially into the past and 
whose time dependence must be included.  It is also assumed that  
$H$ and $\Pi^{IJ}$ which appear in the expression for $\Sigma(\tau_1,\tau_2)$ are 
also sufficiently slowly varying that their time dependence can be neglected.  
The integral is dominated by its upper limit, and these assumptions mean that 
when evaluating 
it one takes $\Pi^{IJ} \to G^{IJ}(\tau_*)$ and $H \to H(\tau_*)$. The assumptions need only be true 
for a short period around the time the initial conditions are fixed.
In the resulting expressions for the 
initial conditions for $B^{abc}$, we keep both the terms which grow fastest as $\tau \to -\infty$ as well as the sub-leading terms. 
The results are rather long to present, and so are given in appendix~\ref{apdxA} together with some further details of the 
calculation. 

We note that all the initial conditions are the simply covariant versions of those for the canonical 
case presented in Ref.~\cite{Dias:2016rjq} with no new terms appearing (except through the extra Riemann terms in the $a$ and $b$ tensors).

\section{The curvature perturbation}
\label{section5}
Thus far we have discussed the framework in which the power spectrum and bispectrum 
of covariant field perturbations can be calculated. These are however not directly related to observations. A quantity often used to make the connection between primordial 
perturbations and observational constraints is the curvature perturbation on 
uniform density slices, $\zeta$.

To calculate the statistics of $\zeta$ we need to know how it is related to 
the set of perturbations $\{Q^I, P_J\}$. 
We require only the form of this relation on super-horizon scales, and we write it in the form
\begin{equation}
\zeta({\bf{k}})=N_{\bf{a}}\delta X^{\bf{a}} + \frac{1}{2}N_{{\bf{ab}}}\delta X^{\bf{a}} \delta X^{\bf{b}}\,,
\end{equation}
where
\begin{equation}
\begin{split}
N_{\bf{a}}({\bf{k}})=&(2\pi)^3\delta({\bf{k}}-{\bf{k_a}})N_a\\
N_{\bf{ab}}({\bf{k,k_a,k_b}})=&(2\pi)^3\delta({\bf{k}}-{\bf{k_a}}-{\bf{k_b}})N_{ab}({\bf{k_a,k_b}})\,.
\end{split}
\end{equation}
In this notation the two 
and the three-point function of $\zeta$ are given 
by
\begin{equation}
\begin{split}
\langle\zeta({\bf{k_1}})\zeta({\bf{k_2}})\rangle =& (2\pi)^3\delta({\bf{k_1}}+{\bf{k_2}})P(k)\\
\langle\zeta({\bf{k_1}})\zeta({\bf{k_2}})\zeta({\bf{k_3}})\rangle=& (2\pi)^3\delta({\bf{k_1}}+{\bf{k_2}}+{\bf{k_3}})B(k_1,k_2,k_3)\,,
\end{split}
\end{equation}
with
\begin{equation}
\begin{split}
\label{psbBispec}
P(k)=&N_aN_b\Sigma^{ab}_{\text{\tiny Re}}(k)\\
B(k_1,k_2,k_3)=& N_a N_b N_c B^{abc}(k_1,k_2,k_3)+ (N_a N_b N_{cb}({\bf{k_1}},{\bf{k_2}})\Sigma^{ac}_{\text{\tiny Re}}(k_1)\Sigma^{bd}_{\text{\tiny Re}}(k_2)+ 2\, cyc.).
\end{split}
\end{equation}

For the case of multi-field inflation 
with canonical kinetic terms, $N_a$ and $N_{ab}$ were 
calculated in Ref.~\cite{Dias:2014msa} (also see Refs.~\cite{Christopherson:2014bea,Carrilho:2015cma}).
Here we extend the calculation 
to the case of a non-trivial field-space metric. 

A first step in the calculation of $\zeta$ in terms of field-space fluctuations on a flat hypersurface is to 
relate $\zeta$ to the total density perturbation on the flat hypersurface. This calculation was performed 
in Ref.~\cite{Dias:2014msa}, and is unchanged in our new setting. One finds
\be
\label{zetatot}
\zeta = - H\frac{\delta \rho}{\dot{\rho}}+H \frac{\dot{\delta \rho}\delta \rho}{\dot{\rho}^2} - \frac{H}{2}\frac{\ddot{\rho}\delta\rho^2}{\dot{\rho}^3} + \frac{\dot{H}}{2}\frac{\delta\rho^2}{\dot{\rho}^2}\,.
\ee

\subsection{The density perturbation}

The new element for the
non-trivial field-space case
is therefore to calculate $\delta \rho$ in this setting. In general, one finds that $\rho =- T^{00}/g^{00}$
\cite{Malik:2008im}, 
where $T_{\mu \nu}$ is the energy momentum tensor. The perturbation in the density up to second order is therefore
\begin{equation}
\delta\rho = \delta T^{00}+\rho\delta g^{00}+ \left(\delta T^{00} +\rho \delta g^{00}\right)\delta g^{00}\,.
\label{drho}
\end{equation} 
For  an arbitrary number of scalar fields 
with non-trivial field-space metric the energy momentum tensor is given by
\begin{equation}
\label{SE}
T_{\mu\nu}=G_{IJ}\partial_{\mu}\phi^I\partial_{\nu}\phi^J - \frac{1}{2}G_{IJ}g_{\mu\nu}\partial^{\lambda}\phi^I\partial_{\lambda}\phi^J-g_{\mu\nu}V.
\end{equation}
This leads to the background energy density $\rho=\frac{1}{2}G_{IJ}\dot{\phi}^I\dot{\phi}^J+V$ 
as expected. Perturbing Eq.~(\ref{SE}) and using Eq.~(\ref{drho}) and recalling that 
\begin{equation}
\begin{split}
g^{00}+\delta g^{00} &= -1 +2\Phi_1 + 2\Phi_2 - 3\Phi_1^2\\
g^{0i}+\delta g^{0i} &= \partial^i\theta_1 +\partial^i\theta_2 - 2\Phi_1\partial^i\theta_1\\
g^{ij}+\delta g^{ij} &= h^{ij} -\partial^i\theta_1\partial^j\theta_1\,,
\end{split}
\end{equation}
one finds that
\begin{equation}
\begin{split}
\delta \rho =& \frac{1}{2}G_{IJ}(\dot{\phi}^I\dot{\delta\phi}^J + \dot{\phi}^J\dot{\delta\phi}^I) - \Phi_1G_{IJ}(\dot{\phi}^I\dot{\delta\phi}^J + \dot{\phi}^J\dot{\delta\phi}^I) +\frac{1}{2}\delta G_{IJ}(\dot{\phi}^I\dot{\delta\phi}^J + \dot{\phi}^J\dot{\delta\phi}^I) \\
& +\frac{1}{2}G_{IJ}\dot{\delta\phi}^I\dot{\delta\phi}^J - \Phi_1G_{IJ}\dot{\phi}^I\dot{\phi}^J + \frac{1}{2}(3\Phi_1-2\Phi_2)G_{IJ}\dot{\phi}^I\dot{\phi}^J + \frac{1}{2}\delta G_{IJ}\dot{\phi}^I\dot{\phi}^J \\ & - \Phi_1\delta G_{IJ}\dot{\phi}^I\dot{\phi}^J + V_{;I}\delta\phi^I + \frac{1}{2}V_{;(IJ)}\delta\phi^I {\delta\phi}^J\,.
\end{split}
\end{equation}

Finally, we need to rewrite this expression in terms of the covariant perturbations, 
$Q^I$ instead of the 
raw field perturbations $\delta \phi^I$. Collecting 
some terms together and applying the relations (\ref{QdPhi}), (\ref{dQdPhi}) and (\ref{dGdPhi})
we obtain a neat expression which at linear order gives
\begin{equation}
\label{edpert1}
\delta\rho_{1} = - \Phi_1G_{IJ}\dot{\phi}^I\dot{\phi}^J+G_{(IJ)}\dot{\phi}^ID_tQ^J+ V_{;I}Q^I\,,
\end{equation}
and at second order 
\begin{equation}
\label{deltaphiR}
\begin{split}
\delta\rho_{2} =& \frac{1}{2}R_{L(IJ)M}\dot{\phi}^L\dot{\phi}^MQ^IQ^J + \frac{1}{2}V_{;(IJ)}Q^IQ^J - 2\Phi_1 G_{(IJ)}\dot{\phi}^ID_tQ^J \\ & + \frac{1}{2}G_{IJ}\dot{\phi}^I\dot{\phi}^J(3\Phi^2_1-2\Phi_2)+\frac{1}{2}G_{IJ}D_tQ^ID_tQ^J.
\end{split}
\end{equation}
Moreover, one can use Eqs.~(\ref{P1}) and (\ref{P2}) to substitute for $\Phi_1$ and $\Phi_2$ and write $\delta \rho$ entirely in terms of the covariant perturbations $Q^I$. 
There are in fact a number of equivalent ways to write $\delta \rho$ as a function of the field-space perturbations using Eq.~(\ref{ham1}) and (\ref{constraint2}), 
which on substitution into Eq.~(\ref{zetatot}) lead to equivalent ways to write $\zeta$ 
in terms of $Q^I$. Different possibilities were discussed at length in Ref.~\cite{Dias:2014msa} for the canonical case. For the numerical implementations of Ref.~\cite{Dias:2016rjq} the simplest of these was used, 
which follows from the use of Eq.~(\ref{P1}) and (\ref{P2}), 
and in the non-trivial field-space case leads to
\begin{equation}
\delta\rho_{1} = - 3HG_{IJ}\dot{\phi}^IQ^J \,,
\label{drhoF1}
\end{equation}
and 
\begin{equation}
\begin{split}
\delta\rho_{2} =&\,\, 3M_\mathrm{p}^2H^2(3\Phi_1^2-2\Phi_2)  \\
=&\,\, \frac{3}{2M_\mathrm{p}^2}\dot{\phi}_I\dot{\phi}_JQ^IQ^J - 3H\partial^{-2}\left(G_{IJ}(\partial_iD_t Q^I)\partial^iQ^J+G_{IJ}D_tQ^I\partial^2Q^J\right).
\end{split}
\label{drhoF2}
\end{equation}

\subsection{The $N$ tensors}

Substituting Eqs.~(\ref{drhoF1}) and (\ref{drhoF2}) into Eq.~(\ref{zetatot}) one finds 
\be
\zeta_{(1)}=-\frac{1}{2M_\mathrm{p}^2H\epsilon}G_{IJ}\dot{\phi}^IQ^J\,,
\ee
and
\begin{equation}
\begin{split}
\zeta_{(2)}=&\frac{1}{6M_\mathrm{p}^2H^2\epsilon}\left[\left(\frac{1}{M_\mathrm{p}^2}\dot{\phi}_I\dot{\phi}_J\left[-\frac{3}{2}+\frac{9}{2\epsilon}+\frac{3}{4\epsilon^2M_\mathrm{p}^2H^3}V_{;K}\dot{\phi}^K\right]\right)Q^IQ^J\right. \\[1ex] &\left.
+\left( \frac{3}{M_\mathrm{p}^2H\epsilon}\dot{\phi}_I\dot{\phi}_J\right)Q^ID_tQ^J - 3H\partial^{-2}\left(G_{IJ}(\partial_iD_tQ^I)\partial^iQ^J + G_{IJ}(D_tQ^I)\partial^2Q^J\right)\right].
\end{split}
\end{equation}
On moving to Fourier space we can identify expressions for the 
$N$ tensors defined above, and we find that
\begin{equation}
N_a = -\frac{1}{2M_\mathrm{p}^2H\epsilon}\dot{\phi}_I
\left(\begin{matrix}
1
\\[1ex]
0
\end{matrix}
\right)
\end{equation}
\begin{equation}
N_{ab} = -\frac{1}{3M_\mathrm{p}^2H^2\epsilon}
\left(\begin{matrix}
\frac{1}{M_\mathrm{p}^2}\dot{\phi}_I\dot{\phi}_J\left[-\frac{3}{2}+\frac{9}{2\epsilon}+\frac{3}{4\epsilon^2M_\mathrm{p}^2H^3}V_{;K}\dot{\phi}^K\right] & \frac{3}{H\epsilon}\frac{\dot{\phi}_I\dot{\phi}_J}{M_\mathrm{p}^2}-G_{IJ}\frac{3H}{k^2}\left({\bf{k_a}}\cdot {\bf{k_b}} + k_a^2 \right)
\\
\frac{3}{H\epsilon}\frac{\dot{\phi}_I\dot{\phi}_J}{M_\mathrm{p}^2}-G_{IJ}\frac{3H}{k^2}\left({\bf{k_a}}\cdot {\bf{k_b}} + k_b^2 \right)& 0
\end{matrix}
\right)\,.
\end{equation}
We note that these equations are simply the covariant from of the canonical case presented 
in Ref.~\cite{Dias:2014msa} with no new terms appearing. It should be noted, however, that 
additional Riemann terms do appear in intermediate expressions, for example for $\delta \rho$ (\ref{deltaphiR}).

\section{Numerical implementation}
\label{Sectionwithstuff}
\label{numerics}

\subsection{PyTransport\,2.0}
So far in this paper we have developed the theoretical framework necessary 
to perform a numerical evolution of the power spectrum and bispectrum 
for models of inflation with a non-Euclidean field-space metric. Now we turn to 
their practical application. 

The equations presented have been implemented in 
an new version of the open source  \texttt{PyTransport} \cite{Mulryne:2016mzv} 
package, \texttt{PyTransport\,2.0}. 
To use this package, an end user is required 
to specify the model they wish to analyse
(in terms of the potential and the field-space), then the code compiles a 
bespoke python module which contains functions that enable the user to calculate 
the evolution of the background fields, 
the evolution of the covariant field-space correlations, and the power-spectrum and bispectrum 
of $\zeta$. The code is released at \href{https://github.com/ds283/CppTransport}{github.com/ds283/CppTransport} with accompanying user manual explaining in detail the 
steps needed to set up the package and apply it to models of interest. 
\subsection{Applications to models of inflation}
\label{appl}
To demonstrate the utility of our framework and  numerical implementation, here we present results we 
have generated  
for a number of models. 

In order to illustrate these 
numerical results we define some quantities that are useful when studying a model of inflation. 
The dimensionless power spectrum, $\mathcal{P}$, of the curvature perturbations, $\zeta$, is defined by
\begin{equation}
\mathcal{P}(k)=\frac{k^3}{2\pi^2}P(k),
\end{equation}
where $P(k)$ was defined in Eq.~(\ref{psbBispec}), and the reduced bispectrum of $\zeta$ by
\begin{equation}
\frac{6}{5}f_{\rm nl}(k_1,k_2,k_3)=\frac{B(k_1,k_2,k_3)}{P(k_1)P(k_2)+P(k_1)P(k_3)+P(k_2)P(k_3)}\,.
\end{equation}
For one triangle of wavevectors in 
the bispectrum, it is often convenient to 
use a parameter to describe the overall scale, $k_s=k_1+k_2+k_3$, 
and two further parameters for the shape, $\alpha$ and $\beta$, defined as
\begin{equation}
\begin{split}
k_1&= \frac{k_s}{4}(1+\alpha+\beta)\\
k_2&=\frac{k_s}{4}(1-\alpha + \beta)\\
k_3&=\frac{k_s}{2}(1-\beta),
\end{split}
\end{equation}
with the allowed values of $(\alpha, \beta)$ falling inside a triangle in the $\alpha$, $\beta$ plane  with 
vertices $(-1,0)$, $(1, 0)$ and $(0, 1)$.

\subsection{ Model with a continuous curved trajectory}
\label{curved}

\begin{figure}
\centering
\begin{subfigure}{.5\textwidth}
  \centering
  \includegraphics[width=1\linewidth]{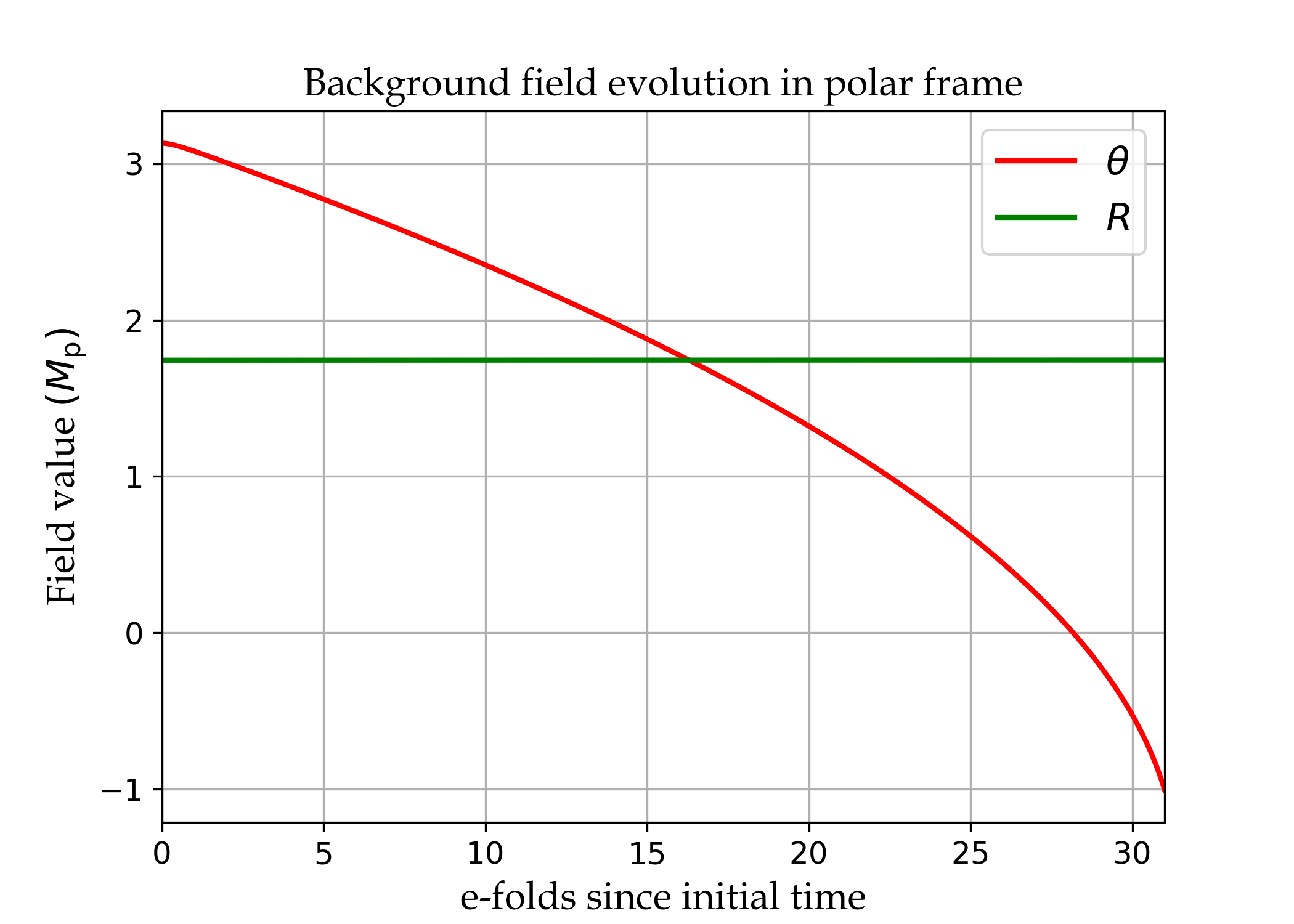}
  \caption{}
  \label{SPHBKEVO}
\end{subfigure}%
\begin{subfigure}{.5\textwidth}
  \centering
  \includegraphics[width=1\linewidth]{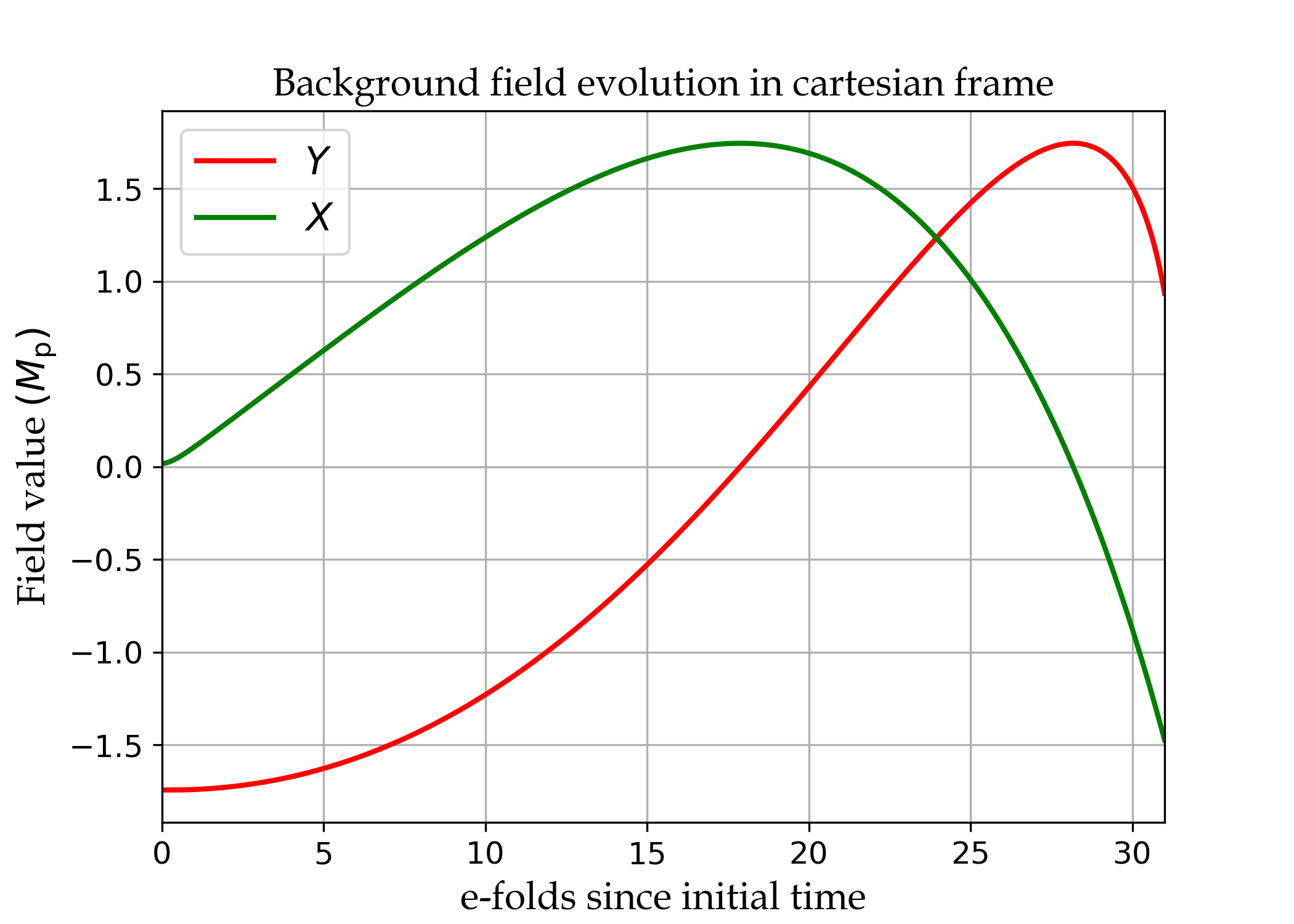}
  \caption{}
  \label{BKEVO}
\end{subfigure}
\caption{The time evolution of the polar coordinate fields $\theta$ and $R$ with metric (\ref{polarmet}) on the left, and the cartesian coordinates, $X$ and $Y$ on the right.}
\label{fig:test}
\end{figure}
\begin{figure}
\centering
\begin{subfigure}{.5\textwidth}
  \centering
  \includegraphics[width=1\linewidth]{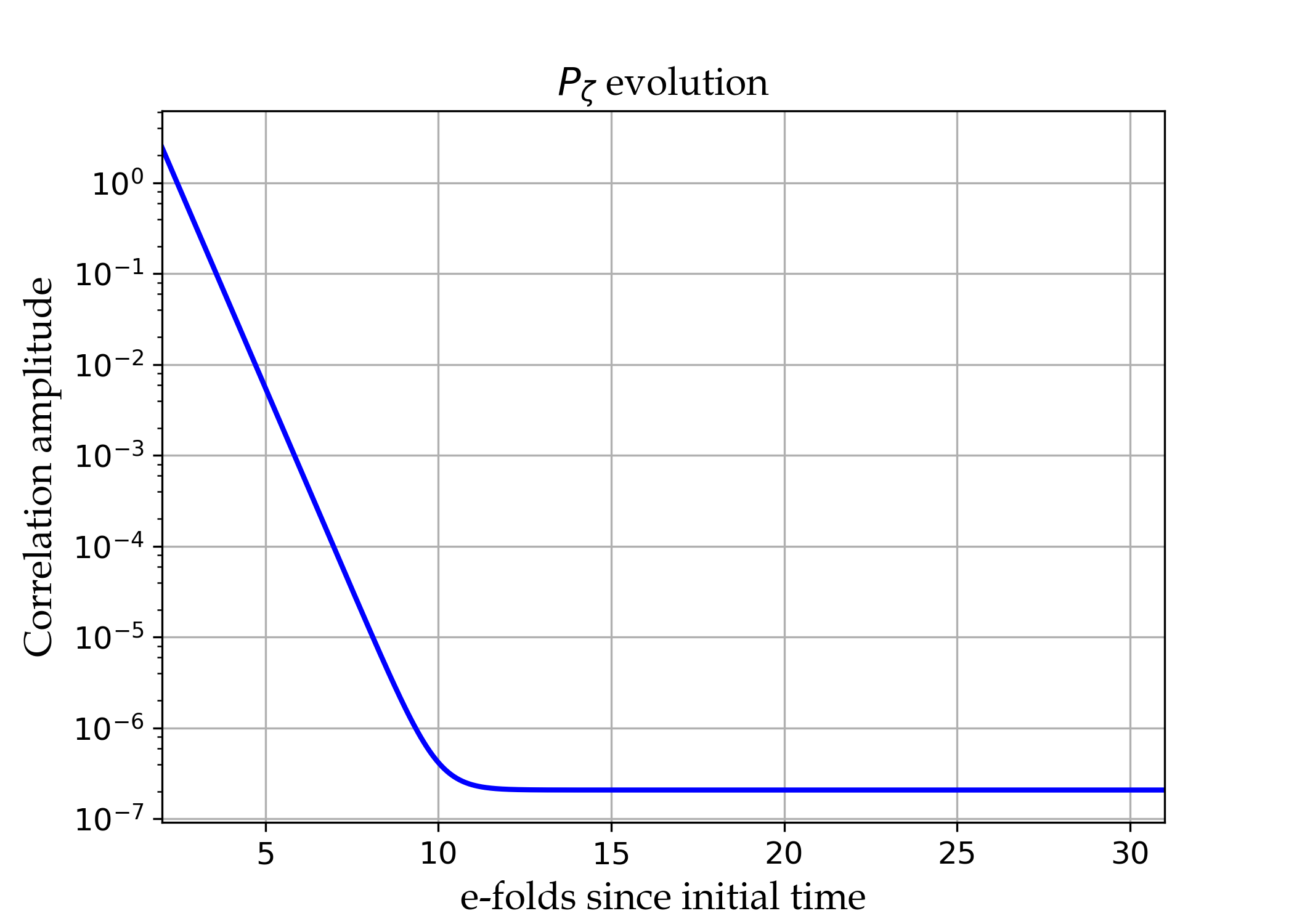}
  \caption{}
  \label{ZETAEVO}
\end{subfigure}%
\begin{subfigure}{.5\textwidth}
  \centering
  \includegraphics[width=1\linewidth]{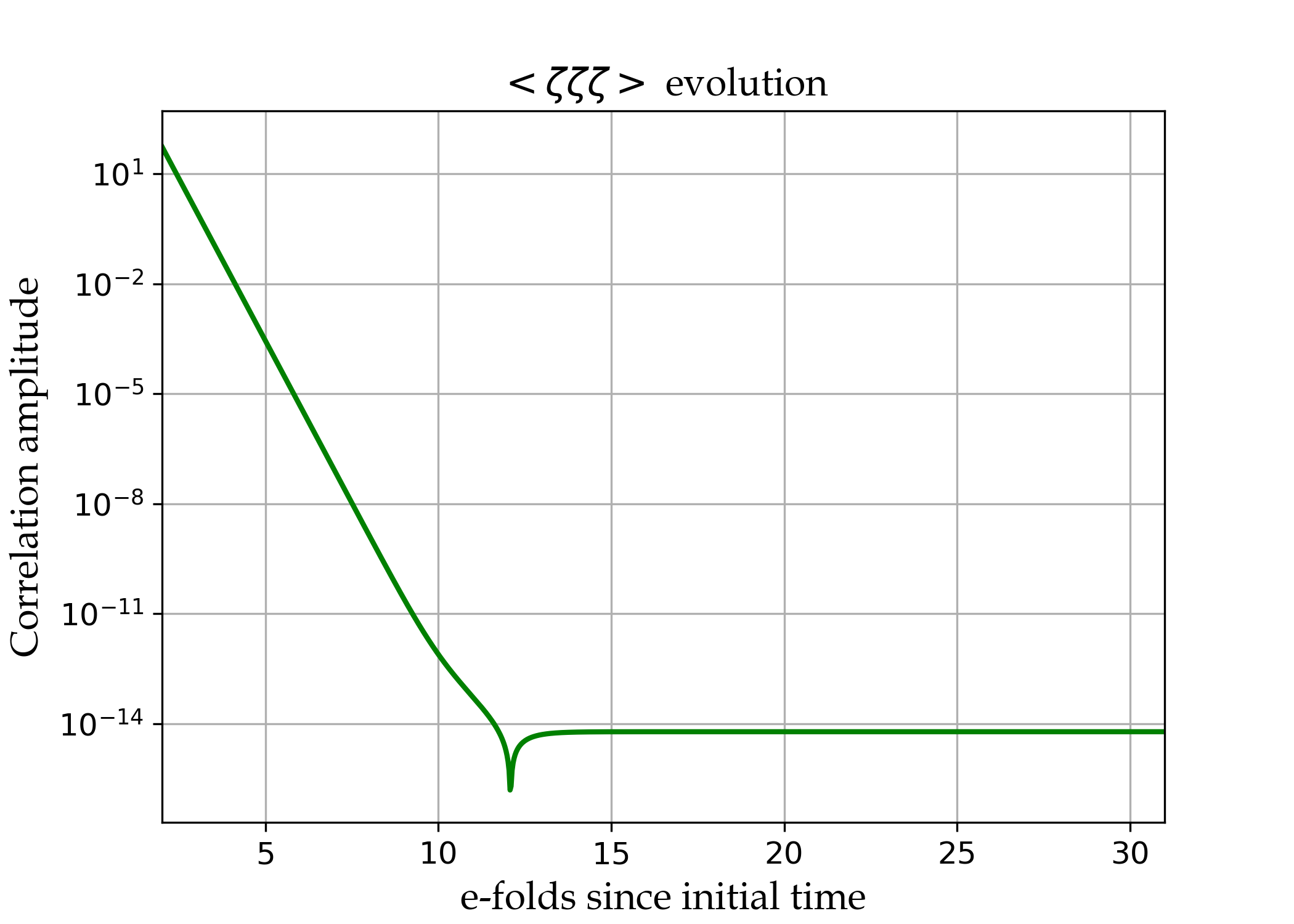}
  \caption{}
  \label{ZZZEVO}
\end{subfigure}
\caption{The time evolution of correlation functions. On the left the time evolution of the two-point function of the curvature perturbation, $\zeta$, and on the right the evolution of the three-point function for an equilateral configuration. Both were taken for modes exiting the horizon 21 e-folds before the end of inflation.}
\label{fig:test}
\end{figure}
\begin{figure}
\centering
\begin{subfigure}{.5\textwidth}
  \centering
  \includegraphics[width=1\linewidth]{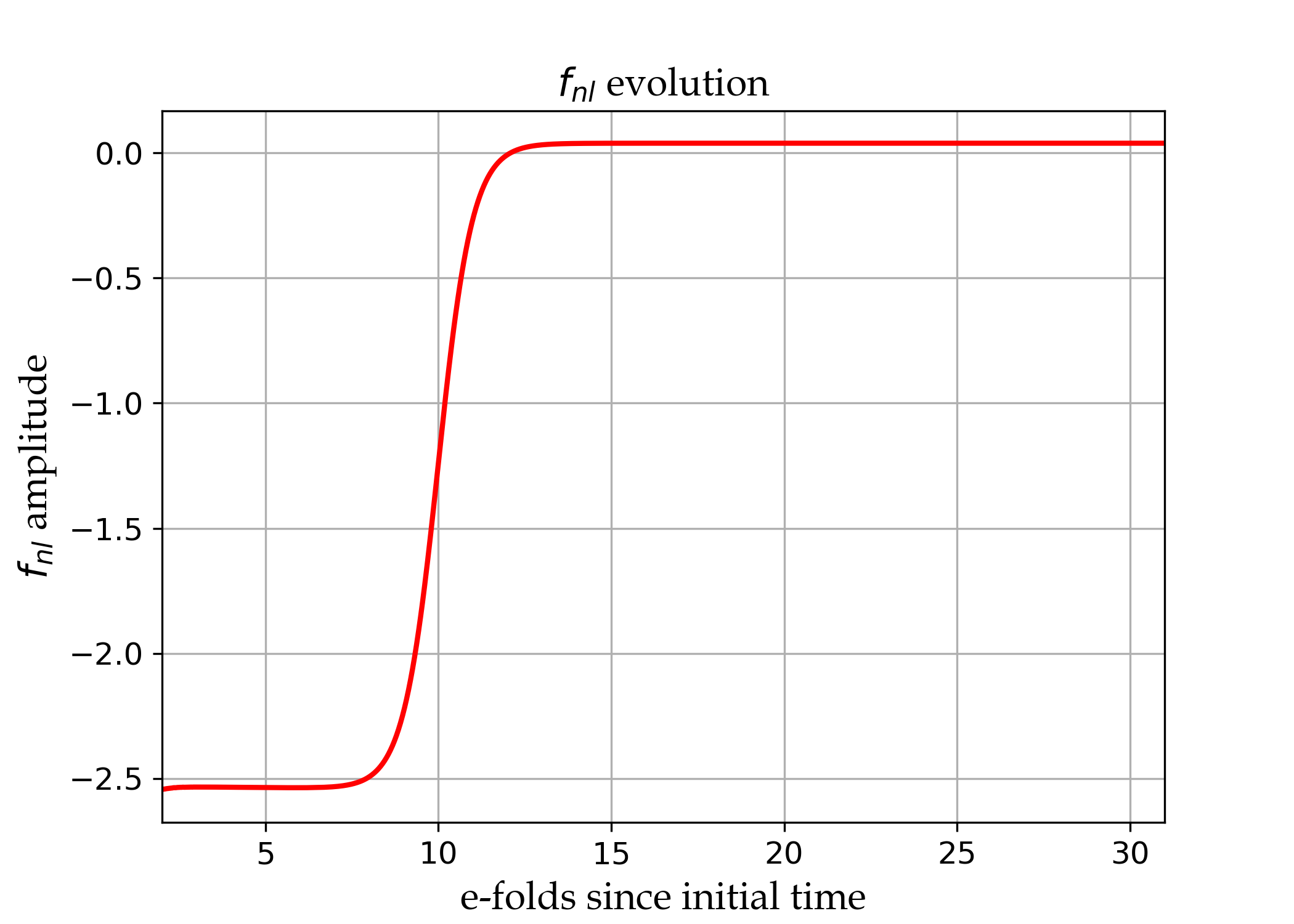}
  \caption{}
  \label{FNL}
\end{subfigure}%
\begin{subfigure}{.5\textwidth}
  \centering
  \includegraphics[width=1\linewidth]{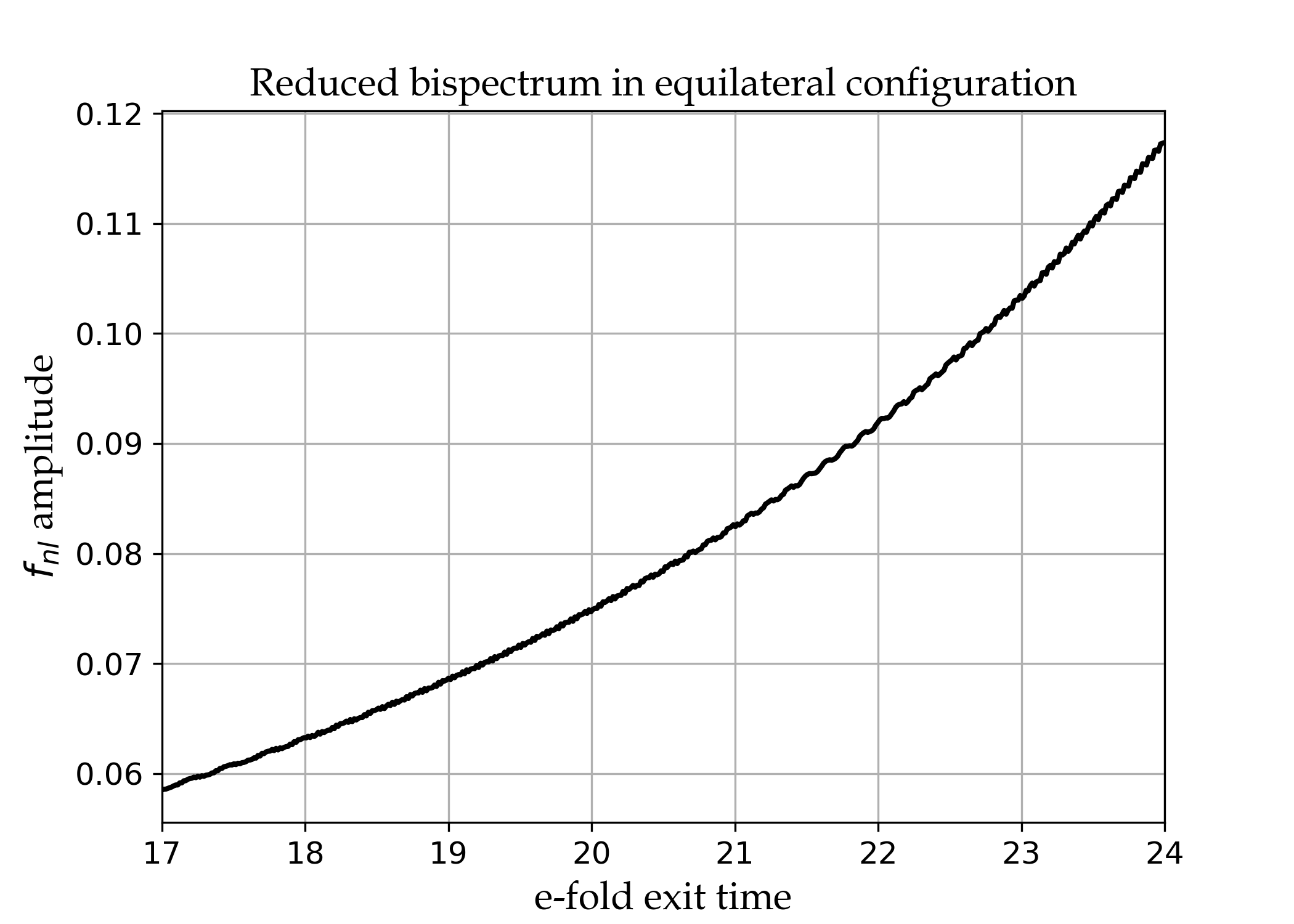}
  \caption{}
  \label{BiEq2}
\end{subfigure}
\caption{The reduced bispectrum $f_{nl}(k_1,k_2,k_3)$ for equilateral configurations. On the left the evolution of $f_{nl}$ versus time for an equilateral configuration with modes leaving the horizon 21 e-folds prior to the end of inflation. On the right the bispectrum over a range of equilateral configurations as a function of exit time of the scale $k_s/3$.}
\label{fig:test}
\end{figure}

Ref.~\cite{Dias:2016rjq} attempted to construct a model in which 
the field-space trajectory was curved in such as way as to 
exhibit Gelaton \cite{Tolley:2009fg} or QSFI \cite{Chen:2009we} behaviour.
For reasons presented there, this behaviour was difficult to achieve, but the 
model presented there is still a useful example, and in the present context 
provides a useful check of our code. 

The model is defined by the action for two fields $R$ and $\theta$ as
\begin{equation}
S= -\frac{1}{2}\int d^4 x\sqrt{-g}\left[(\partial R)^2 + R^2 (\partial \theta)^2 + 2V(R,\theta)\right]\,,
\label{Curve1}
\end{equation}
where the potential (defined below in Eq.~(\ref{curvepot1}))
represents a circular valley at a fixed value of $R$ -- and hence is naturally written in terms of 
these `polar coordinate' fields.
However, as the codes 
developed for Ref.~\cite{Dias:2016rjq} only dealt 
with canonical kinetic terms, in that work it was necessary to perform a field redefinition to cartesian coordinates $X$ and $Y$.
Here we evolve the statistics directly for the fields $R$ and $\theta$ and compare results, 
using this as a test case to benchmark our code against its canonical precursor.

The field-space metric of the model can be read off from Eq.~(\ref{Curve1}), and is 
\begin{equation}
\label{polarmet}
G_{IJ}=\left(
\begin{matrix}
1 & 0\\
0 & R^2
\end{matrix}\right)\,.
\end{equation}
The potential is
\begin{equation}
\label{curvepot1}
V=V_0\left(1+\frac{29\pi}{120}\theta + \frac{1}{2} \frac{\eta_\mathrm{R}}{M_\mathrm{p}^2}(R-R_0)^2+\frac{1}{3!}\frac{g_\mathrm{R}}{M_\mathrm{p}^3}(R-R_0)^3 +\frac{1}{4!}\frac{\lambda_\mathrm{R}}{M_\mathrm{p}^3}(R-R_0)^4 \right)\,,
\end{equation}
and we choose parameters $V_0=10^{-10}M_\mathrm{p}^4$, $\eta_\mathrm{R} = 1/\sqrt{3}$, $g_\mathrm{R}=M_\mathrm{p}^2V_0^{-1/2}$, $\omega=\pi/30$, $\lambda_\mathrm{R}=0.5M_\mathrm{p}^3\omega^{-1/2}V_0^{-3/4}$ and $R_0=\frac{30\sqrt{10^{-10}/3}}{\pi\sqrt{10^{-9}}}$.
With these choices, the radial direction represents a heavy mode confining the 
inflationary trajectory to the valley, with angular direction light. We further choose
initial conditions
\begin{equation}
R_{\mathrm{ini}}=\sqrt{R_0^2+(10^{-2}R_0)^2} \quad {\rm and} \quad \theta_{\mathrm{ini}}=\arctan \left(\frac{10^{-2}R_0}{R_0}\right)\,.
\end{equation}

Generating results using our new code for the field evolution and correlations in the 
$\{R,\theta\}$ basis, and then subsequently using a coordinate transformation to  translate the 
results to the $\{X,Y\}$ basis, we can compare our results to the output of the 
canonical code. We find excellent agreement. The evolution of correlation functions 
of the curvature perturbation, $\zeta$, are coordinate invariant, and also match that generated 
using the canonical code.
In Fig.~\ref{SPHBKEVO} the background field evolution in the non-canonical case is plotted. Under the coordinate transformation to
 the canonical fields $X$ and $Y$ we get the evolution in Fig. \ref{BKEVO}.  In Fig. \ref{ZETAEVO} \& \ref{ZZZEVO} 
one can clearly see that after horizon crossing the curvature perturbation freezes in,
 becoming constant on large scales as expected. The evolution of the reduced Bispectrum 
 $f_{nl}$ for one equilateral triangle is shown in Fig. \ref{FNL}. The reduced bispectrum in the equilateral configuration 
 as a function of horizon crossing time 
 is given in Fig. \ref{BiEq2}, and can be compared with Fig.~11 of Ref. \cite{Dias:2016rjq}.

\subsection{ Quasi-two-field inflation}

\begin{figure}
\centering
\begin{subfigure}{.5\textwidth}
  \centering
  \includegraphics[width=1\linewidth]{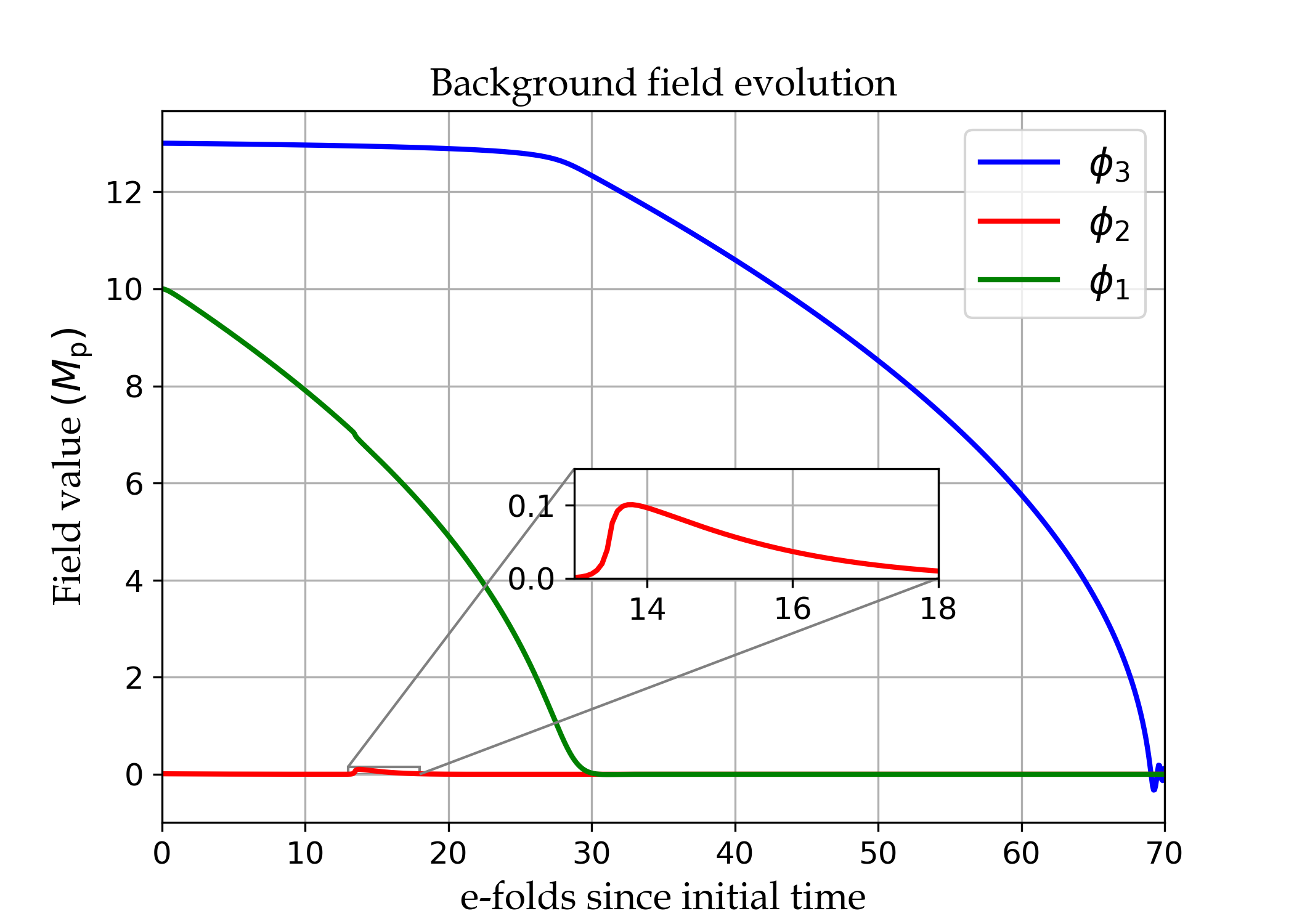}
  \caption{}
  \label{BK2}
\end{subfigure}%
\begin{subfigure}{.5\textwidth}
  \centering
  \includegraphics[width=1\linewidth]{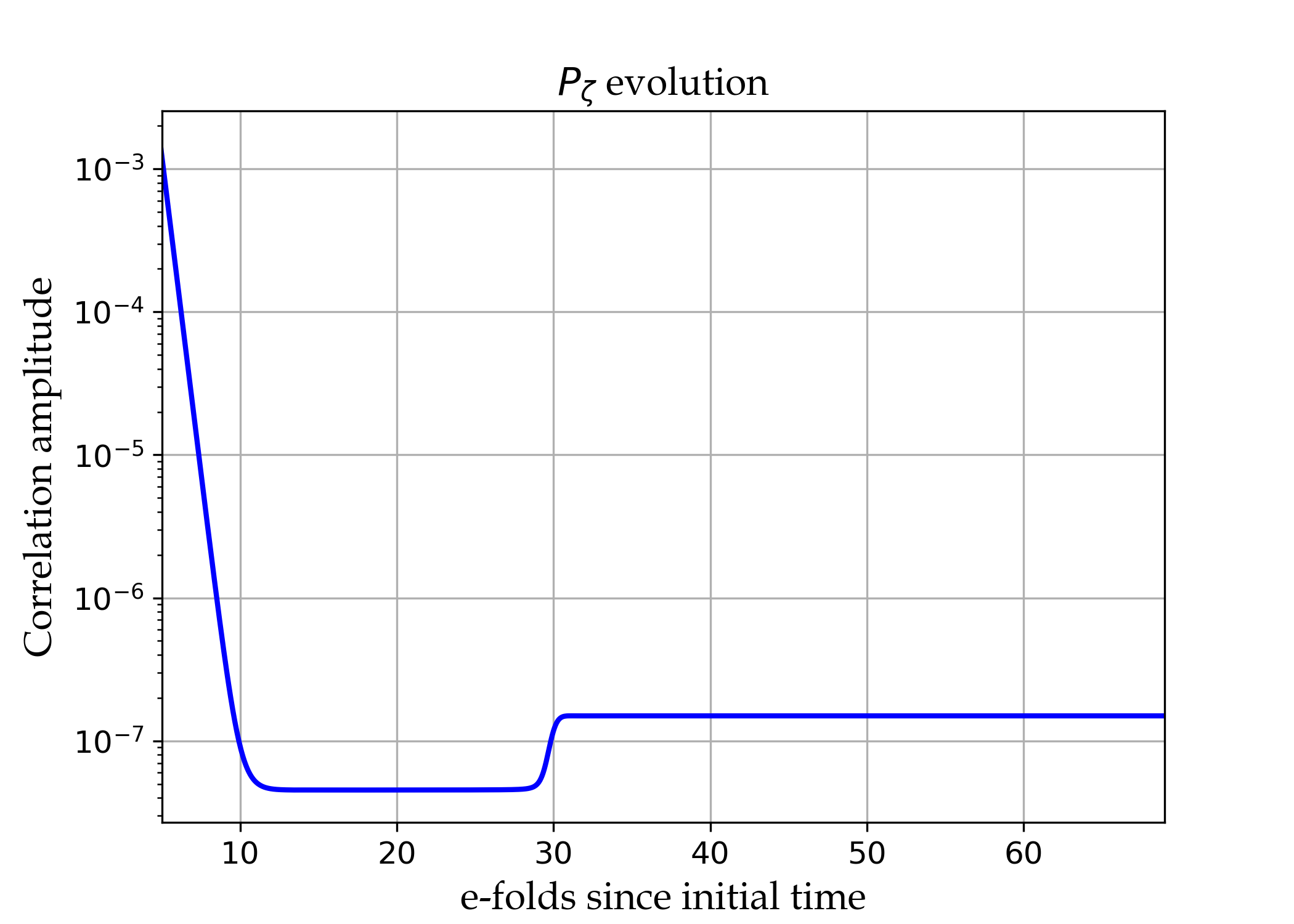}
  \caption{}
  \label{ZZ2}
\end{subfigure}
\caption{The time evolution of the fields $\phi_1$, $\phi_2$ and $\phi_3$ on the left, and the time evolution of the two-point function of $\zeta$ for a k-mode exiting the horizon 60 e-folds before the end of inflation on the right. The turn in field-space occurs 13 e-folds into inflation when the field $\phi_2$ experiences excitations from its coupling to the lighter field $\phi_1$ via the field-space metric. After roughly 30-e-folds the $\phi_1$ field reaches the minimum and the amplitude of the power spectrum increases at this time. }
\label{fig:test}
\end{figure}
\begin{figure}
\centering
  \includegraphics[width=.5\linewidth]{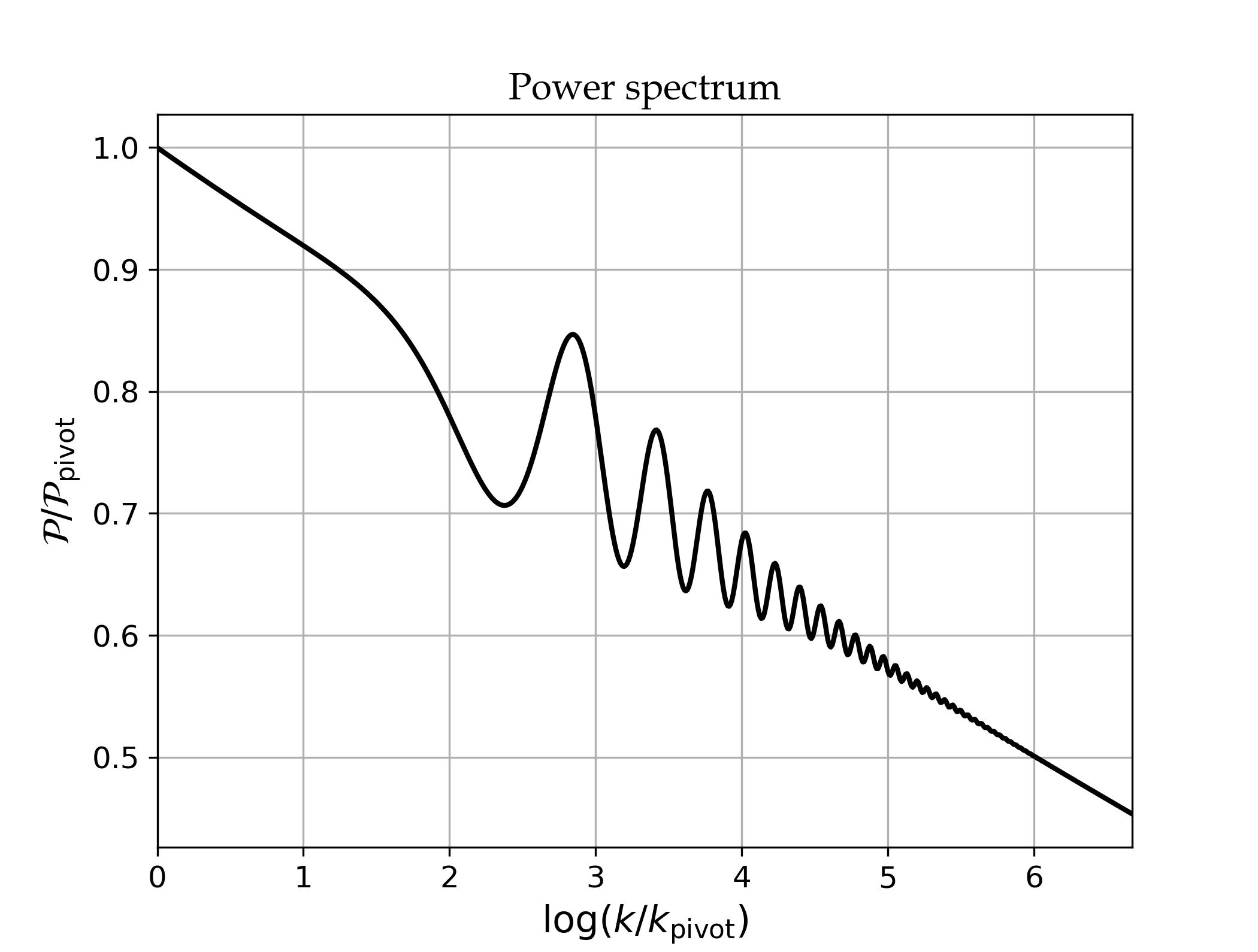}
  \caption{The power-spectrum of the curvature perturbation for a range of modes which exit the horizon over a window of 7 e-folds. The scale $k_{pivot}$ is taken to be when the mode leaves the horizon at 58 e-folds prior to the end of inflation. Both the scales and amplitudes are normalised to the spectrum at the pivot scale.}
\label{PZpseudo2}
\end{figure}
\begin{figure}[h!]
\centering
\begin{subfigure}{.5\textwidth}
  \centering
  \includegraphics[width=1\linewidth]{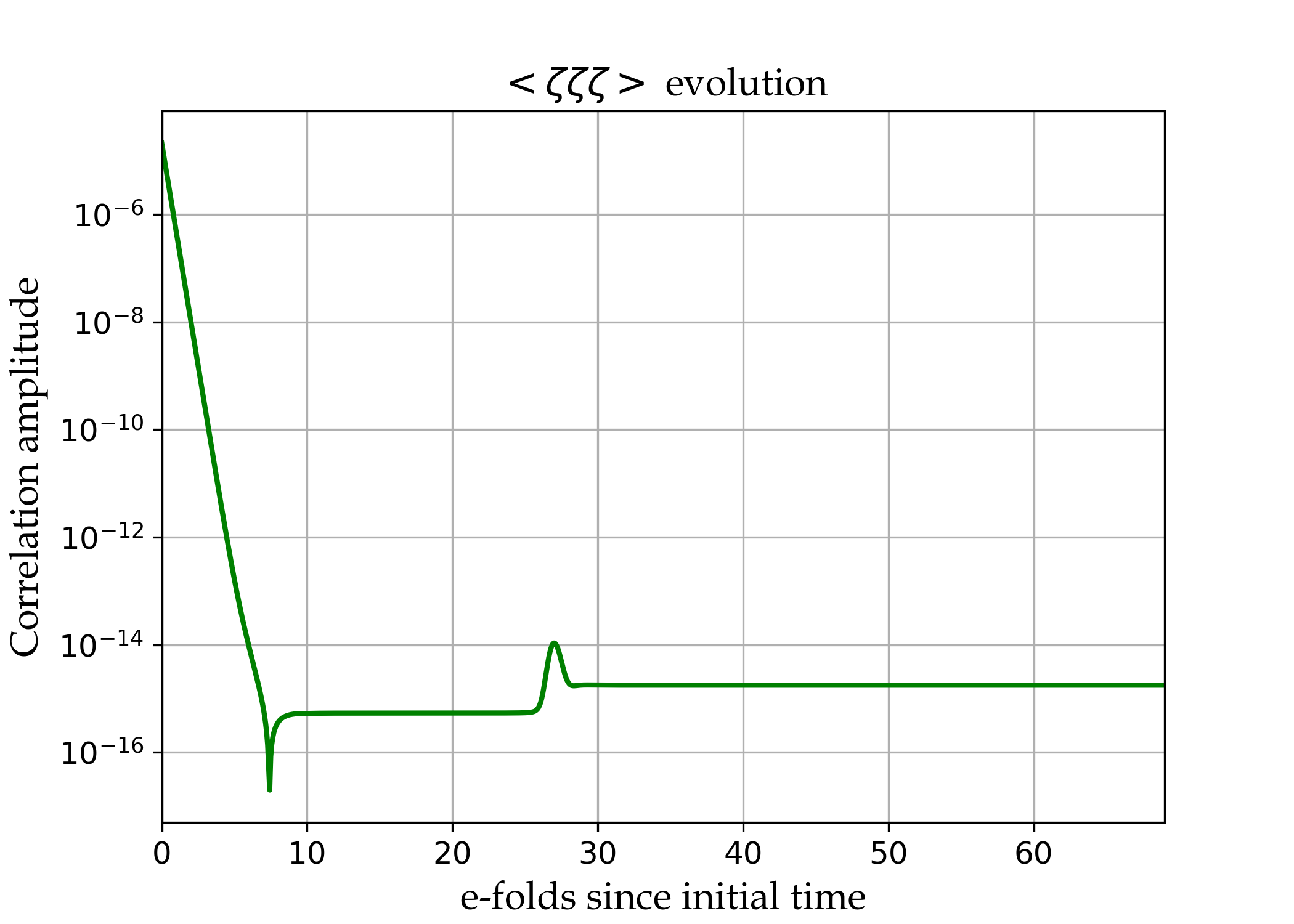}
  \caption{}
  \label{ZZZ3}
\end{subfigure}%
\begin{subfigure}{.5\textwidth}
  \centering
  \includegraphics[width=1\linewidth]{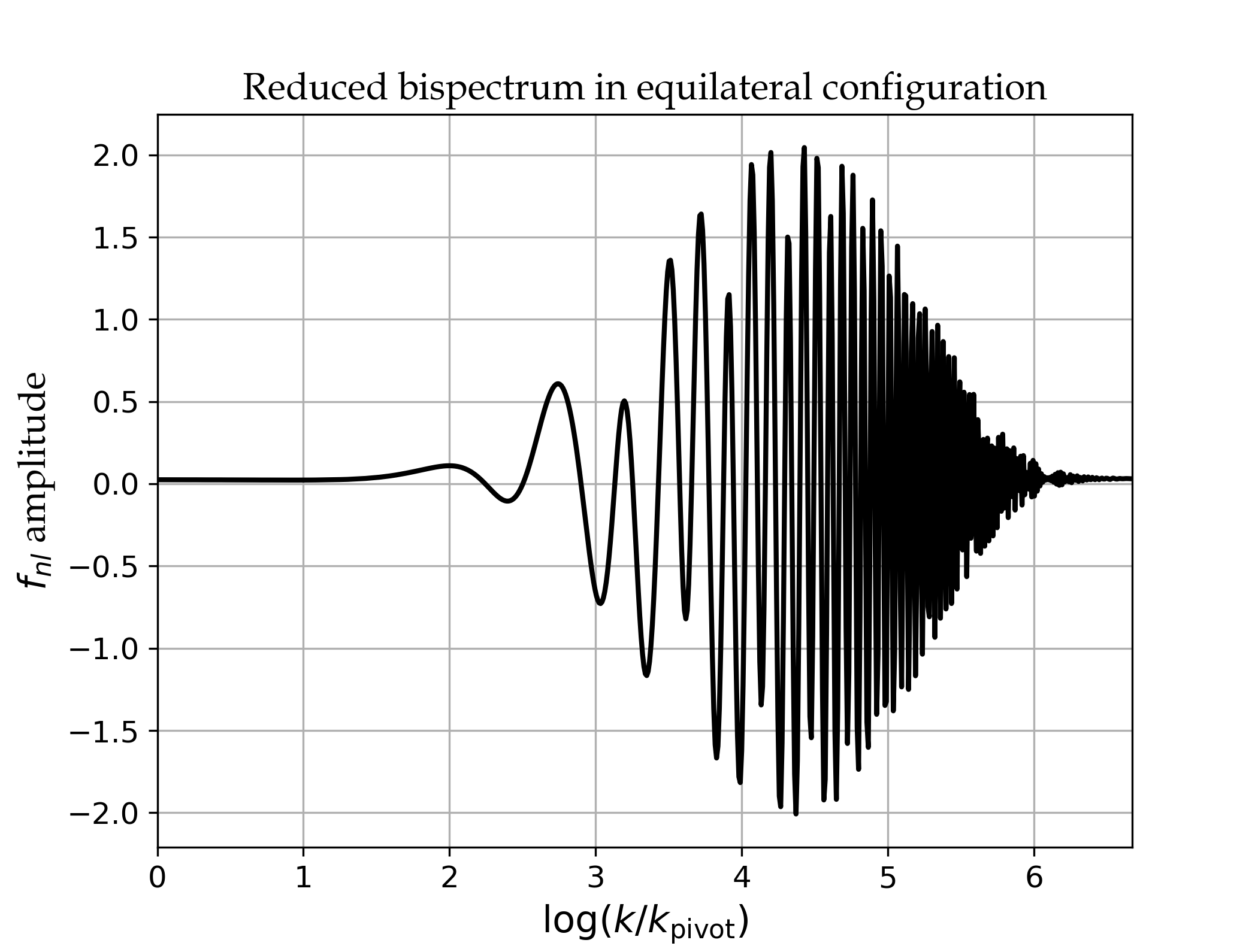}
  \caption{}
  \label{BZpseudo2}
\end{subfigure}
\caption{The evolution of the three-point function for one equilateral configuration, and the reduced bispectrum, $f_{nl}$, 
for equilateral configurations over a range $k_s$. The reduced bispectrum is plotted for modes leaving the horizon between 59 and 51 e-folds before the end of inflation. The highly oscillatory behaviour is a result of the excitations to the heavy field around horizon crossing.}
\label{fig:test}
\end{figure}
\begin{figure}
\centering
  \includegraphics[width=1\linewidth]{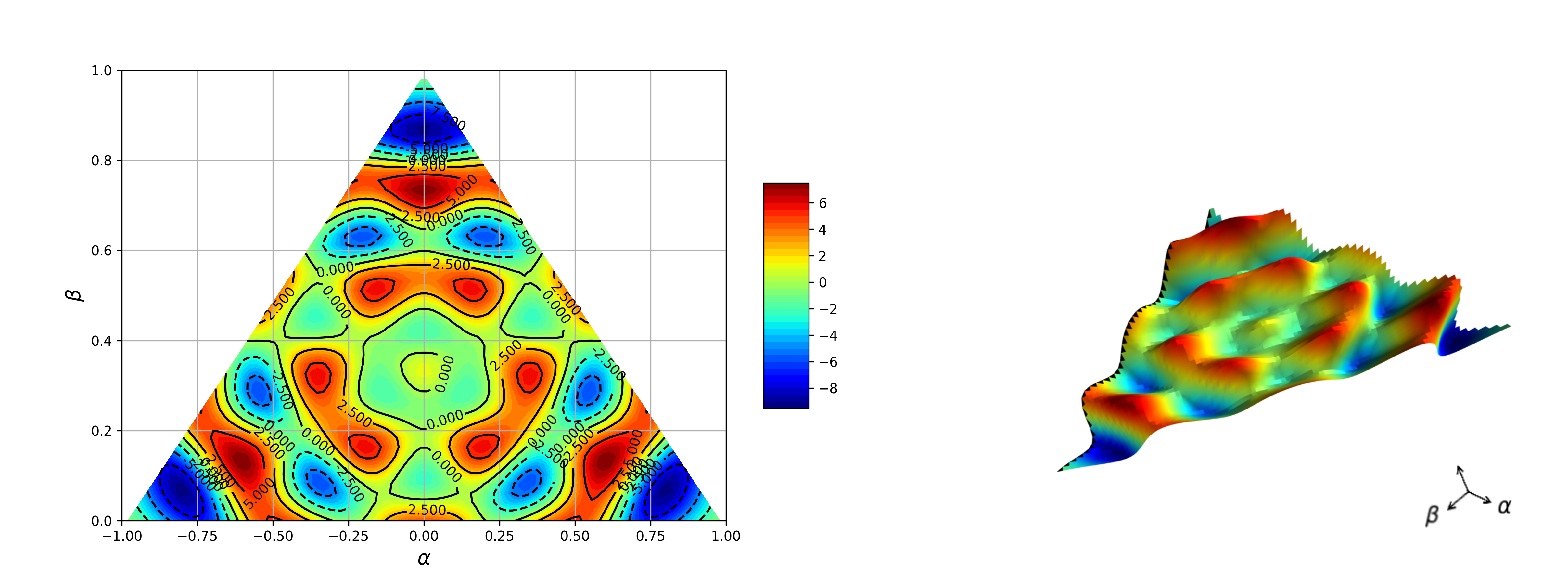}
  \caption{Amplitude over shape configurations of the reduced bispectrum ${f_{nl}}(\alpha,\beta)$ at a fixed $k_t$ 53 e-folds before the end of inflation, corresponding to $\log(k/k_{pivot})=4.79$. }
\label{alpBet}
\end{figure}

Next we consider the 
quasi-two field model introduced in Ref. \cite{Dias:2015rca} where the power spectrum was calculated. 
In this model there are two light scalar fields which drive inflation and one heavy 
field which interacts with the light ones through a coupling in the kinetic terms.
This leads to  a fast turn in the plane of the lighter two fields resulting in the well known feature of oscillations in the power 
spectrum and bispectrum (see for example \cite{Gao:2012uq,Gao:2013ota,Achucarro:2013cva,Achucarro:2014msa,Adshead:2013zfa,Flauger:2016idt,Palma:2014hra,Mooij:2015cxa}). In this paper we reproduce the power spectrum presented in Ref.~\cite{Dias:2015rca}
as a test of our code and then calculate the bispectrum for the first time. 
The three fields are labelled $\phi_1$, $\phi_2$ and $\phi_3$, and 
model has a metric which takes the form
\begin{equation}
G_{IJ}=\left(
\begin{matrix}
1 & \Gamma(\phi_1) & 0\\
\Gamma(\phi_1) & 1 & 0\\
0 & 0 & 1
\end{matrix}
\right)\,.
\end{equation}
The function $\Gamma(\phi_1)$ has the following $\phi_1$ dependence \cite{Achucarro:2010da},
\begin{equation}
\Gamma(\phi_1) = \frac{\Gamma_0}{\cosh^2\left(2\left(\frac{\phi_1-\phi_{1(0)}}{\Delta\phi_1}\right)\right)}\,,
\end{equation}
with $\Gamma_0=0.9$ the maximum value attained by $\Gamma(\phi_1)$. $\phi_{1(0)}=7M_\mathrm{p}$ is the value of $\phi_1$ at the apex of the turn in field-space and $\Delta\phi_1=0.12$ is the range of $\phi_1$ over which the turn occurs
The potential is defined as
\begin{equation}
V= \frac{1}{2}g_{1}m^2\phi_1+\frac{1}{2}g_{2}m^2\phi_2+\frac{1}{2}g_{3}m^2\phi_3\,,
\end{equation}
with parameters $g_{1} =30$, $g_{2}=300$, $g_{3}=30/81$ and $m=10^{-6}$.
The initial conditions of the fields are
\begin{equation}
\phi_1=10.0M_\mathrm{p}\quad \phi_2 = 0.01M_\mathrm{p} \quad \phi_3 = 13.0M_\mathrm{p} \,.
\end{equation}
In Fig. \ref{BK2} the background field evolution is plotted. At 13 e-folds into the evolution the turn in the inflationary trajectory occurs, as can be seen by the increase in the amplitude of the heaviest field. In Figs.~\ref{ZZ2}~\&~\ref{ZZZ3} the evolution of both the two and three-point correlation functions of curvature perturbations are plotted. The power spectrum obtained in Fig.~\ref{PZpseudo2} matches that seen in Ref.~\cite{Dias:2015rca} illustrating that the code is in good agreement with this earlier implementation. We produce the reduced bispectrum over equilateral configurations in Fig.~\ref{BZpseudo2}, the structure of which is defined by a pulse of large and rapidly oscillating values of the three-point function. Finally, for a fixed scale $k_t$ we plot the reduced bispectrum in Fig.~\ref{alpBet} as a function of the $\alpha$ and $\beta$ parameters discussed in \S\ref{appl} for a fixed $k_t$.

\subsection{Inflation on a 2-sphere metric}
\label{2-sphere}

\begin{figure}
\centering
\begin{subfigure}{.5\textwidth}
  \centering
  \includegraphics[width=1\linewidth]{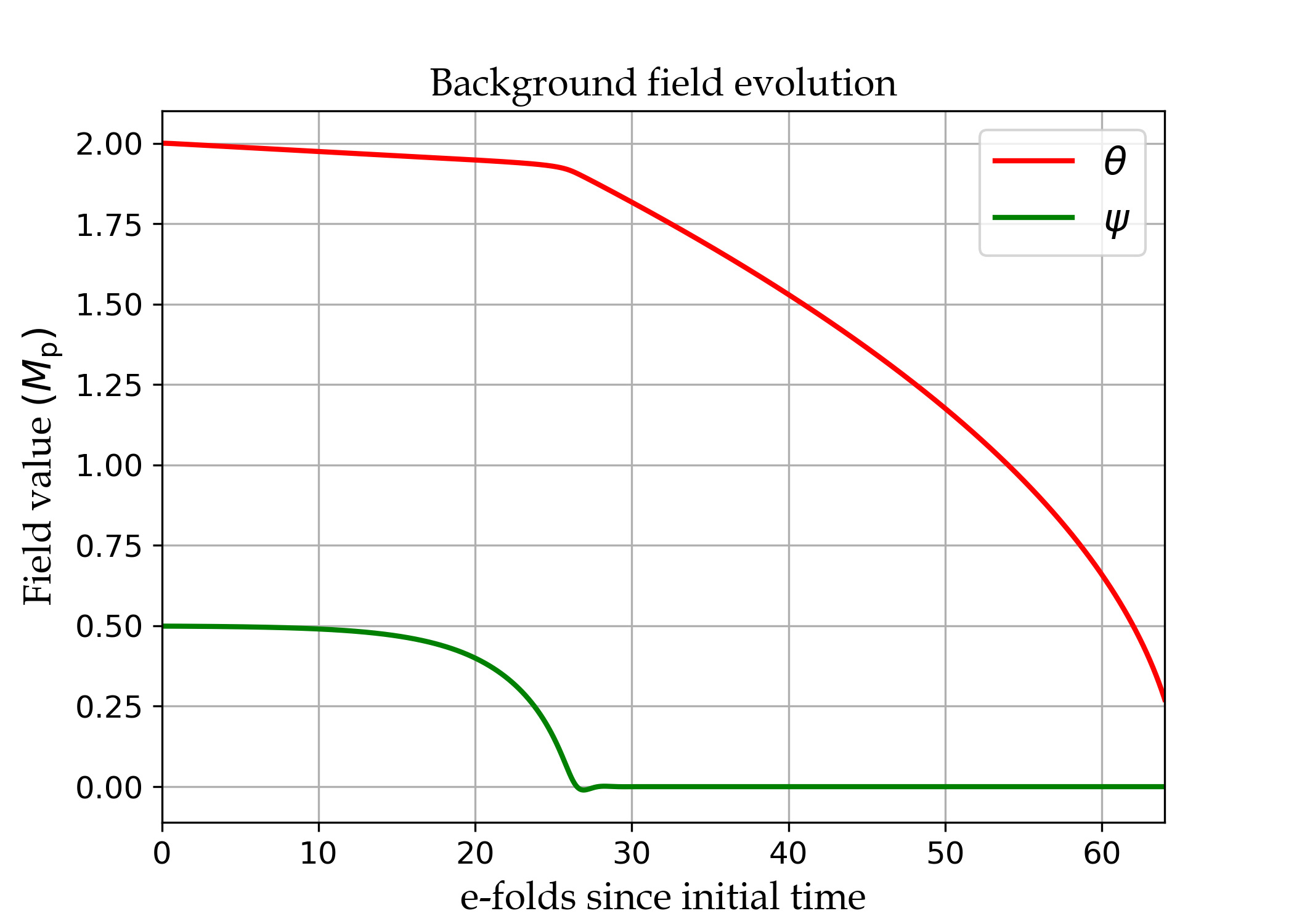}
  \caption{}
  \label{Backevo2SAX}
\end{subfigure}%
\begin{subfigure}{.5\textwidth}
  \centering
  \includegraphics[width=1\linewidth]{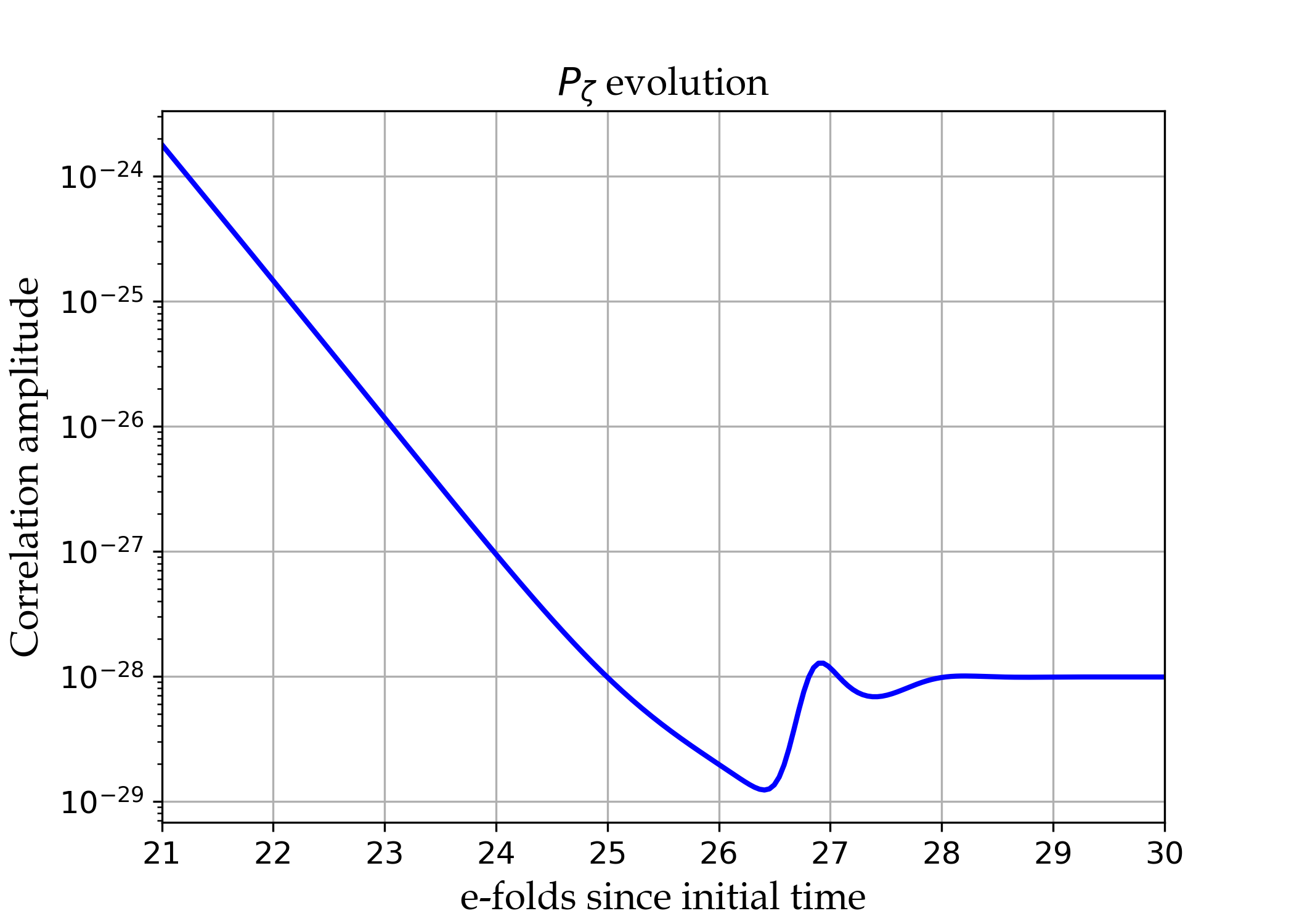}
  \caption{}
  \label{pzevo}
\end{subfigure}
\caption{The evolution of the fields $\theta$ and $\psi$ on the left and the evolution two-point function of  the curvature perturbation 
on the right for a mode leaving the horizon 50 e-folds prior to the end of inflation. 
From 30 e-folds into inflation until the end there is no further evolution of the two-point function.}
\label{fig:test}
\end{figure}
\begin{figure}
\centering
\begin{subfigure}{.5\textwidth}
  \centering
  \includegraphics[width=1\linewidth]{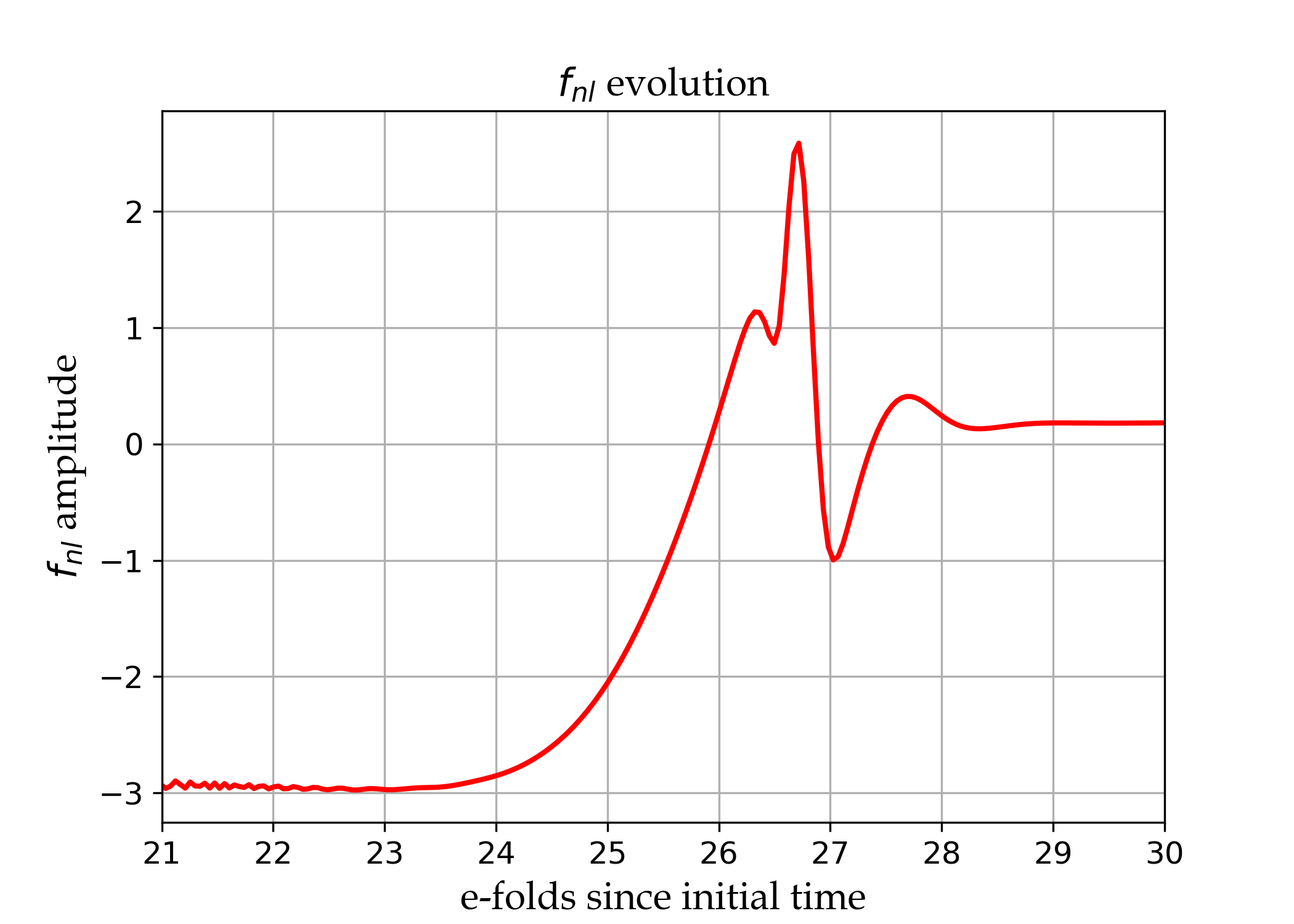}
  \caption{}
  \label{fnlevo2spAX}
\end{subfigure}%
\begin{subfigure}{.5\textwidth}
  \centering
  \includegraphics[width=1\linewidth]{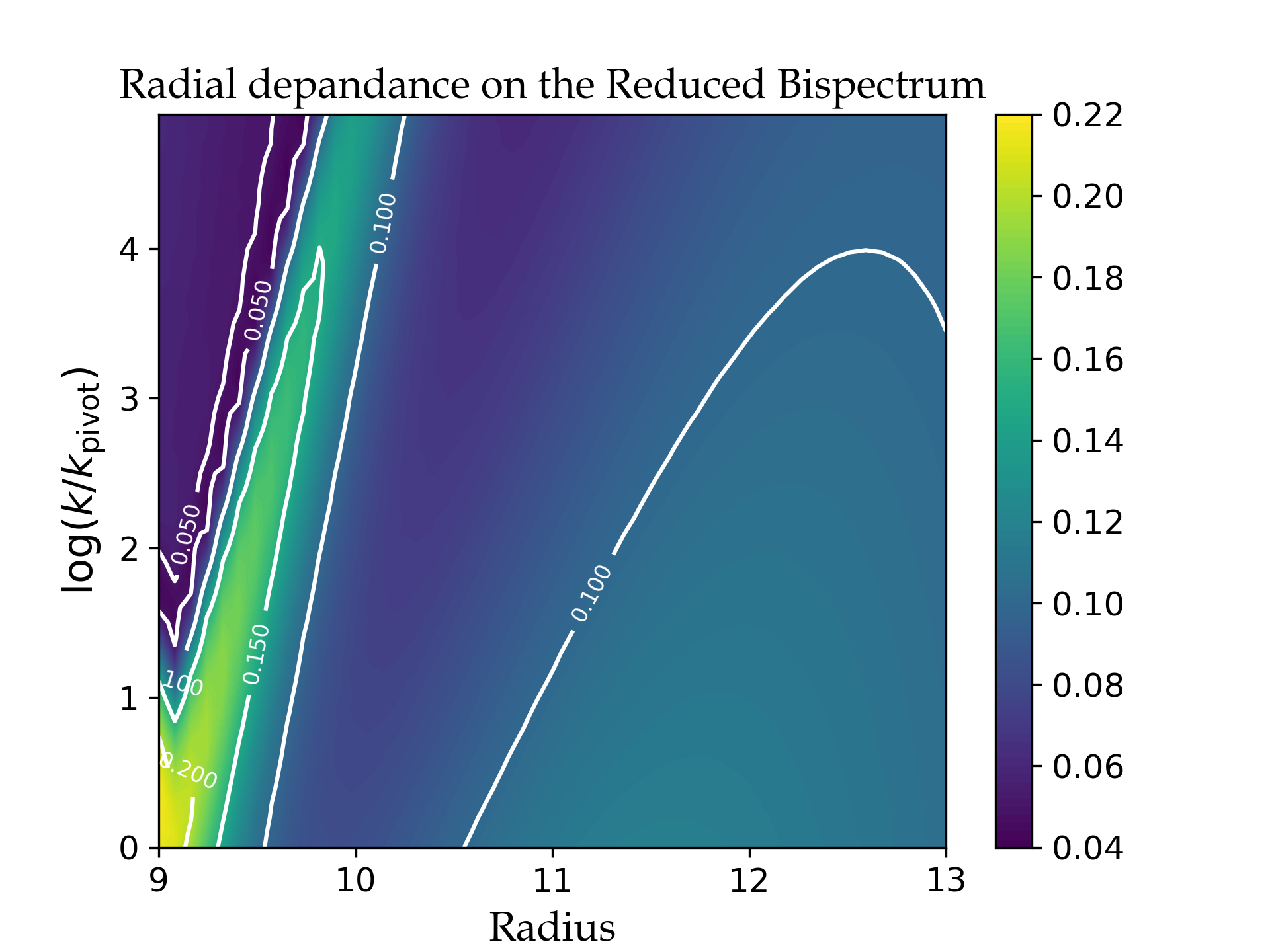}
  \caption{}
  \label{2sphaxBispec}
\end{subfigure}
\caption{Evolution of the reduced bispectrum in an equilateral configuration on the left and the reduced bispectrum for an equilateral configuration versus the radius of the metric sphere on the right. From 30 e-folds into inflation until the end there is no further evolution of $f_{nl}$. The evolution of $f_{nl}$ was taken for a mode leaving the horizon at 26 e-folds from the beginning of inflation. The bispectrum on the right is taken for a range of modes in the window between 25 and 30 e-folds and for a radius between 9 and 11.5. It illustrates a large amplitude correlation over scales for a small radius (or rather large field-space curvature).}
\label{fig:test}
\end{figure}

In the models considered above the field-space metrics were non-trival, but flat.
As a further test of our code, therefore,  
we now introduce a model with a constant non-zero Ricci curvature.

We construct a toy model containing two fields $\theta$ and $\psi$, where the action is defined as
\begin{equation}
S= -\frac{1}{2}\int d^4 x\sqrt{-g}\left[r_0^2(\partial \theta)^2 + r_0^2 \sin^2\theta(\partial \psi)^2 + 2V(\theta,\psi)\right]\,,
\end{equation}
where $r_0$ is the radius of the surface of the sphere which the field trajectory is confined to. The curvature of the field-space, defined by the Ricci Scalar, is related to the radius, $R=\frac{2}{r^2_0}$. 
The field-space metric which describes the line element along the surface of a sphere is therefore
\begin{equation}
G_{IJ}=\left(
\begin{matrix}
r_0^2 & 0\\
0 & r_0^2\sin^2\theta
\end{matrix}\right).
\end{equation}
For the potential we use the same potential given for the  axion-quartic model studied 
in Ref. \cite{Dias:2016rjq}. The potential is of the form,
\begin{equation}
V=\frac{1}{4}g_{\theta}\theta^4 + \Lambda^2\left(1-\cos\left(\frac{2\pi \psi}{f}\right)\right)\,,
\end{equation}
where the field $\psi$ is our ``2-sphere-axion'' and our parameters are $g_{\theta} =10^{-10}$, $\Lambda^4=(25/2\pi)^2gM_\mathrm{p}^4$, $\omega = 30/\pi$ and $f=M_\mathrm{p}$.
The initial conditions of the fields are set to
\begin{equation}
\theta_{\rm ini}=2.0M_\mathrm{p} \quad and \quad \phi_{\rm ini}=f/2-10^{-3}M_\mathrm{p},
\end{equation}
which is sufficient for inflation for 64 e-folds. The background evolution of the fields are plotted in Fig.~\ref{Backevo2SAX}, with the corresponding evolution of correlations of the curvature perturbations for two-point (Fig.~\ref{pzevo}) and three-point (Fig.~\ref{fnlevo2spAX}) functions. 
We study the effects of curvature on quantities like the bispectrum 
by varying the radius $r_0$. Figure ~\ref{2sphaxBispec} is a contour graph of the 
bispectrum as a function of $r_0$. We see that for a radius $r_0>11.0$ the bispectrum is 
small, but for $r_0<11.0$ the bispectrum begins to increase. This indicates a correlation between large curvature and a value of large $f_{nl}$ in this model.

\subsection{Inflation on a conifold metric}

{\begin{figure}
\centering
\includegraphics[width=.5\linewidth]{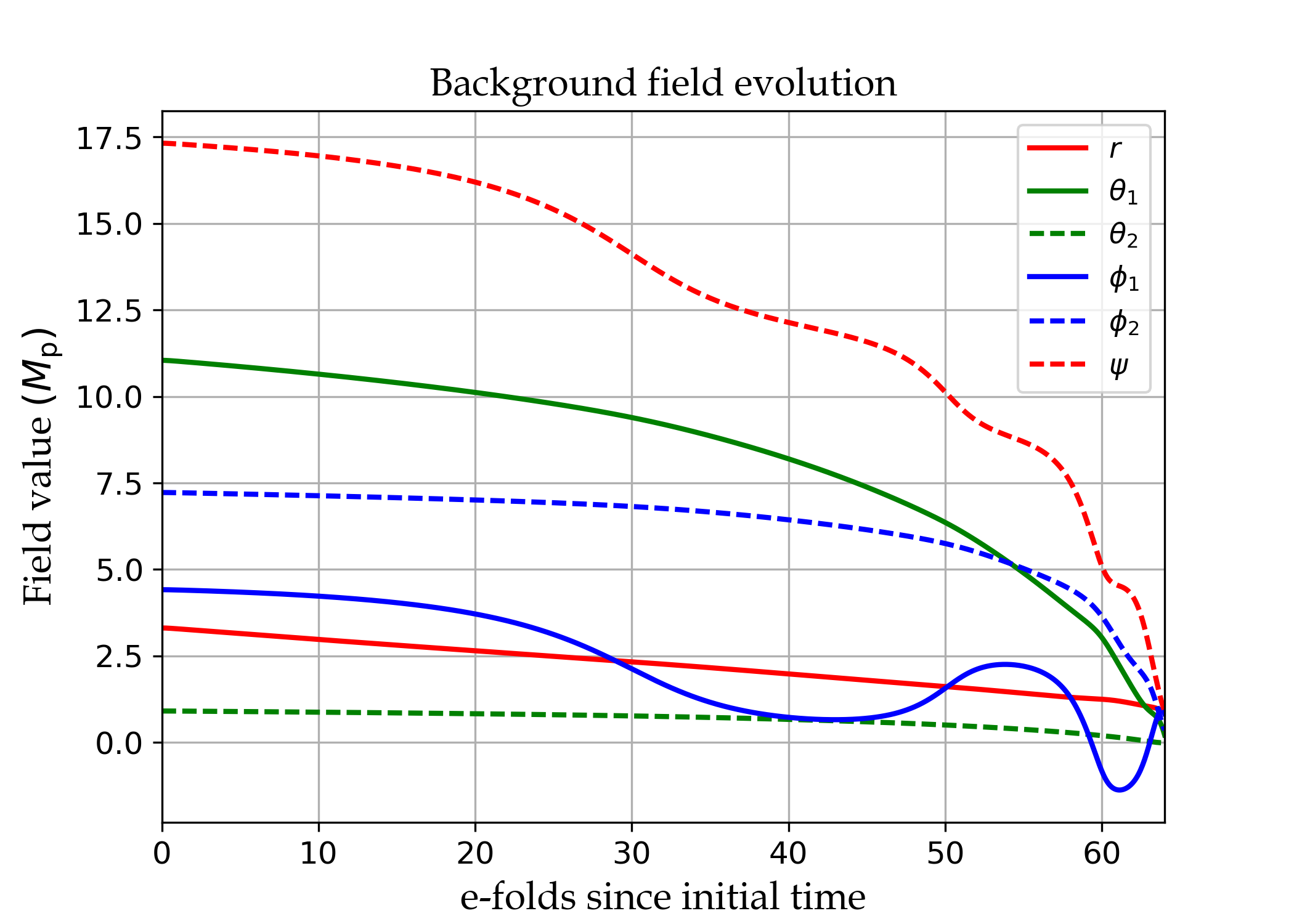}
\caption{The evolution of the 6 moduli fields during inflation. Rich dynamics exist owing to the couplings in the conifold metric. Inflation ends when the branes collide at a value of $r=0$.}
\label{BackevoD6}
\end{figure}
\centering
\begin{figure}[h!]
\begin{subfigure}{.5\textwidth}
  \centering
  \includegraphics[width=1\linewidth]{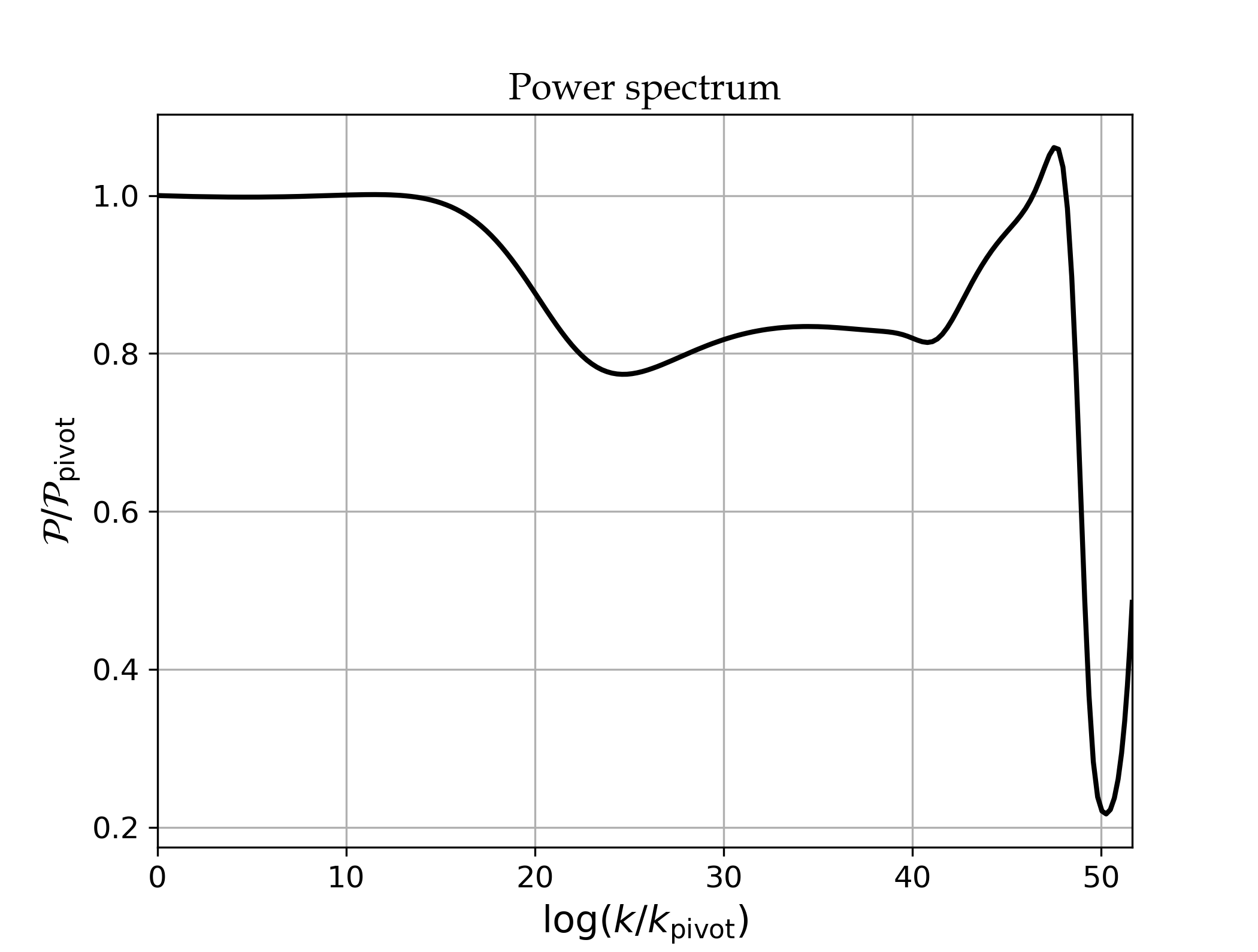}
  \caption{}
  \label{pzD6}
\end{subfigure}%
\begin{subfigure}{.5\textwidth}
  \centering
  \includegraphics[width=1\linewidth]{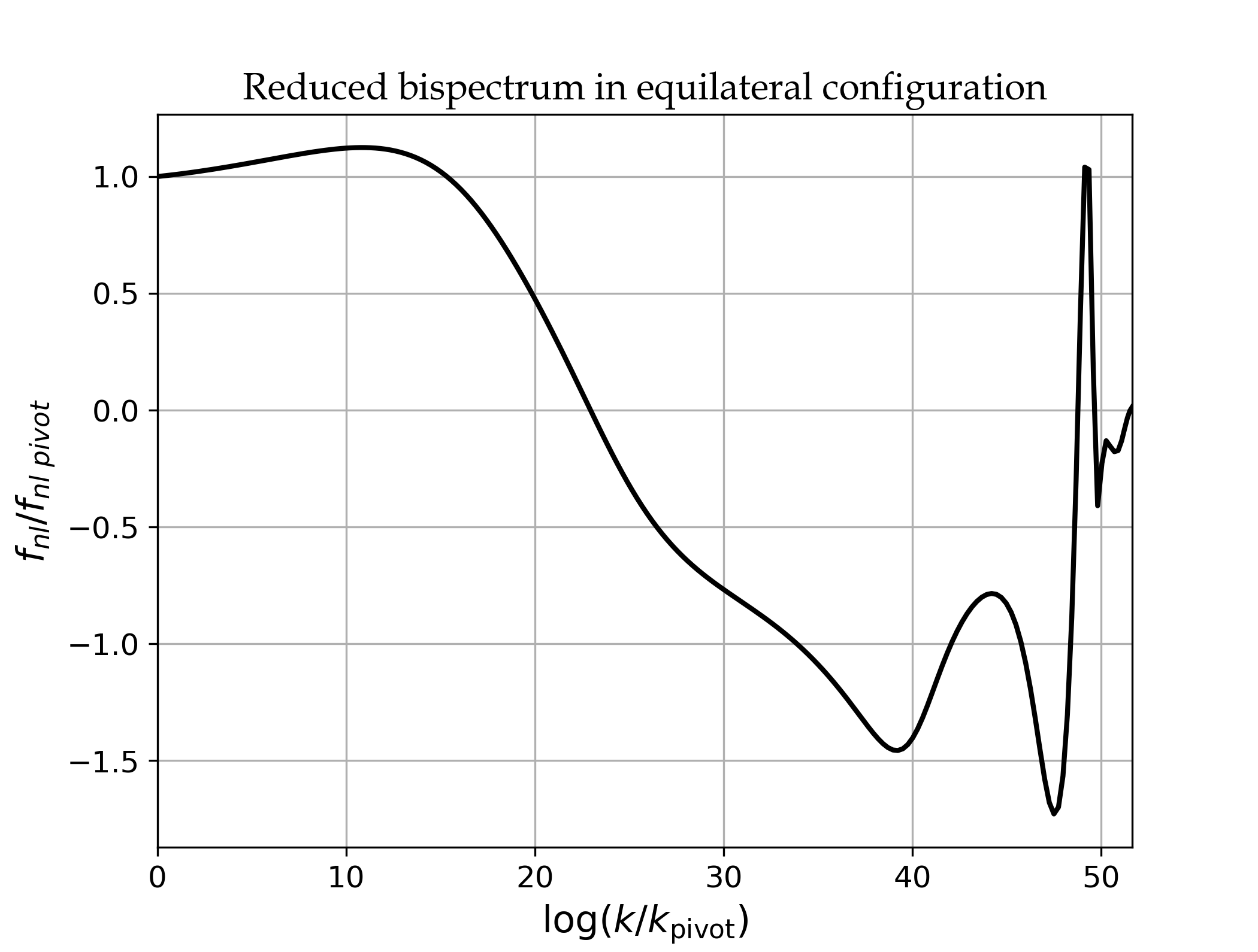}
  \caption{}
  \label{BiEqD6}
\end{subfigure}
\caption{On the left, the power spectrum of curvature perturbation and on the right the bispectrum of curvature perturbations over an equilateral configuration for modes exiting the horizon after a large range of times between 12 and 64 e-folds.}
\label{fig:test}
\end{figure}}

Finally we consider a more realistic case inspired by models of D-brane inflation.
Such models have recently been the subject of considerable interest, with a number of groups statistically 
probing their realisations \cite{Agarwal:2011wm,Dias:2012nf,McAllister:2012am}.
 In one such scenario two D3-branes are attracted by a Coulomb force. 
 Compactification induces a warping of the 6-D manifold where the D3-brane sits, 
 resulting in a non-trivial field-space metric in the Lagrangian of the system. 
 Both the geometry of the metric and structure of the potential affect the inflationary dynamics. 
 Initial work \cite{Agarwal:2011wm} 
 looked at the background dynamics of this system, while more recent studies looked into the distribution 
 of  2-point statistics \cite{Dias:2012nf,McAllister:2012am}. Here we illustrate how our new code could be 
 used to  obtain information about the bispectrum, though we defer realistic studies to future work.
 
We consider the Lagrangian of  D3-brane inflation as
\begin{equation}
S=-\frac{1}{2}\int d^4x\sqrt{-g}\left(G_{IJ}d\phi^Id\phi^J+2V(\phi_1,\dots\phi_6)\right),
\end{equation}
where $a$ is the scale factor. The scalar fields represent the 6 brane coordinates, one radial $r$ and five angular dimensions $\theta_1,\theta_2,\phi_1,\phi_2$ and $\psi$. The field-space metric $G_{IJ}$ corresponds to the Klebanov-Witten conifold geometry \cite{Klebanov:1999tb}. The metric is of the form,
\begin{equation}
G_{IJ}d\phi^Id\phi^J=dr^2+r^2d\Omega^2\,,
\end{equation}
with the metric of the cone $d\Omega$ \cite{Candelas:1989js} is given by
\begin{equation}
d\Omega^2= \frac{1}{6}\sum^2_{i=1}\left(d\theta_i^2+\sin^2\theta_id\phi_i^2\right)+\frac{1}{9}\left(d\psi + \sum^2_{i=1}\cos\theta_id\phi_i\right)^2\,,
\end{equation}
which is a non-compact geometry built over the five-dimensional $(SU(2)\times SU(2))/U(1)$ coset space $T^{1,1}$. As a 
toy example 
we do not generate a realistic potential (motivated by any attractive forces between branes or contribution 
from either the homogeneous or the inhomogeneous bulk), instead, for simplicity, we take a 
quadratic potential for the 6 fields
\begin{equation}
V(\phi)=\sum^6_{i=1}m_i^2\phi_i^2\,,
\end{equation}
where $m_i$ are the randomised masses of the fields. A randomised set of masses and initial conditions are selected with the criteria that 64 e-folds of inflation occur. With these parameters the evolution of the dynamics and statistics can be run and the background trajectory for each of the six fields is plotted in Fig.~\ref{BackevoD6}. The power spectrum is plotted in Fig.~\ref{pzD6} and the bispectrum in the equilateral configuration is plotted in Fig.~\ref{BiEqD6}. 
It would be interesting to run a more realistic analysis including the full potential of the system 
but this is beyond the scope of our work. We have, however, demonstrated that this is 
possible using the transport method and its implementation in code via \texttt{PyTransport}.

\subsection{Performance}

\begin{figure}
\centering
\includegraphics[width=1\linewidth]{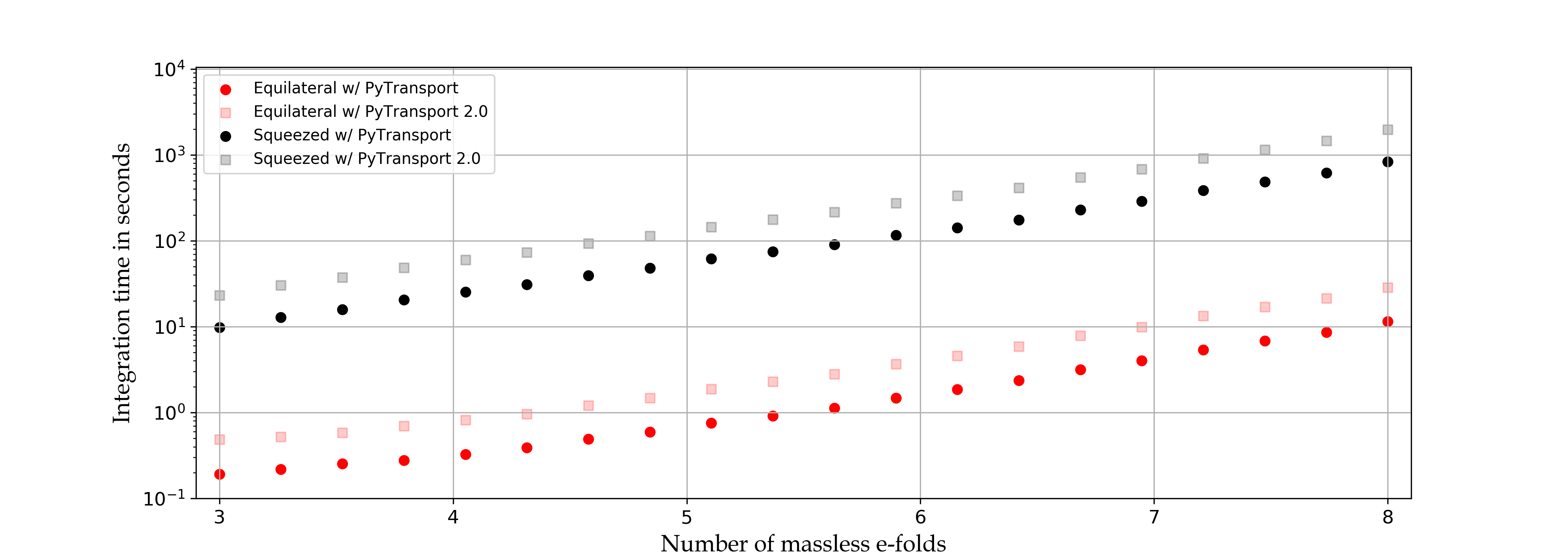}
\includegraphics[width=1\linewidth]{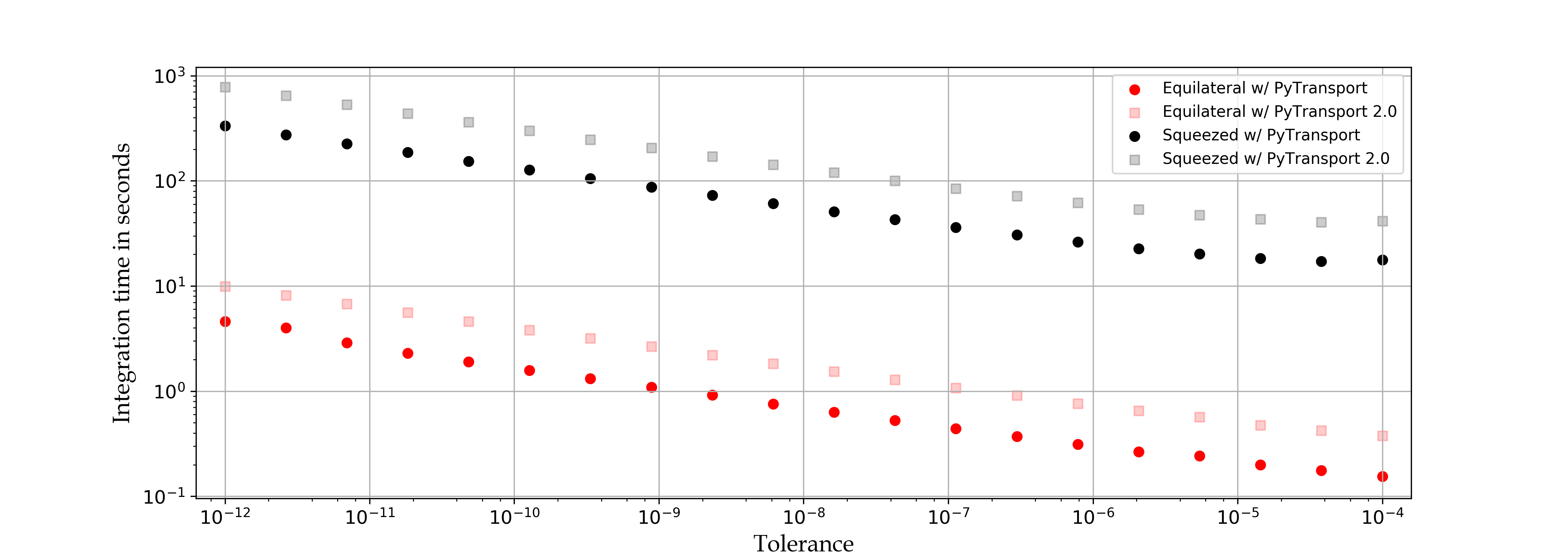}
\caption{Top panel: scaling of integration time with increasing number of massless (or subhorizon) e-folds (using relative and absolute tolerances of $10^{-8}$) for an equilateral triangle of the bispectrum and a squeezed triangle ($\alpha = 0$, $\beta=0.99$). Timings were performed using the canonical code and the new non-canonical code setting a Euclidean metric explicitly.  
Bottom panel: scaling of integration time with integration tolerance with 5 e-folds of massless evolution. The double quadratic model used to analysis performance in Ref.~\cite{Dias:2016rjq} is timed using the canonical PyTransport package and compared to the same model using PyTransport 2.0. The computer used for timings contained an  3.1 GHz Intel i7-4810MQ processor.}
\label{timings}
\end{figure}

In  \texttt{PyTransport\,2.0}, one 
can opt to specify explicitly a field space metric. If this option 
is not selected the code defaults to assuming that the metric is Euclidean and the code reverts back to the 
previous canonical code. The simplicity of a Euclidean 
metric means that a number of internal loops do not need to be performed, and 
hence the canonical code is expected to be faster than when a metric is specified explicitly (even if the 
metric is the Euclidean one). To demonstrate this effect and also to benchmark the speed of the new code 
in Fig.~\ref{timings} we show how the speed of the new code compares with 
that of the canonical one. We also show how the speed of the code is sensitive to the number of e-folds before 
horizon crossing (of the shortest scale in the triangle being evaluated) at 
which initial conditions are fixed, and to different tolerances which fix the accuracy of the code. 
For this purpose we use the double quadratic potential used to calculate performance data in 
Ref.~\cite{Dias:2016rjq}.
As can be seen, the new code is roughly a factor of 2 slower for this two field model. We find that introducing a 
simple field space metric, such as the 2-sphere metric used in \S\,\ref{2-sphere}, leads to 
very similar timing data to the Euclidean metric (though more complicated metrics will inevitably slow 
down the code as the terms in the metric need to be evaluated at each time step). 
A more significant effect comes from increasing the number of fields. 
The size of the arrays which store information about the Riemann tensor and its derivative scale as ${\cal N}^4$ 
and ${\cal N}^5$ respectively (for the canonical code the largest arrays scale as ${\cal N}^3$), and therefore 
memory issues and overheads resulting form accessing and looping over 
these arrays grow rapidly as field number increases.

\section{Conclusion}
\label{conclusions}

We have extended the method of calculating the power spectrum and bispectrum developed in Ref.~\cite{Dias:2016rjq} for canonical multi-field inflation to include models which contain a non-trivial field-space metric.
First in \S\ref{section2} the equations of motion and the conservation equations for perturbations  
were derived for our non-canonical multi-field system. We reviewed how the system of equations 
can be written as an autonomous system for a set of covariant ``field'' perturbations. 
Next we reviewed how the transport method is applied in combination with these equations 
to give equations for the evolution of the correlations of the covariant perturbations during inflation.
To use this system in practice we needed to calculate both initial conditions for our new system of equations, and the 
relation between covariant field-space perturbations and the curvature perturbation $\zeta$. A neat result we found  
is that our expressions for these quantities 
take the form of the covariant versions of the expressions presented in Ref.~\cite{Dias:2016rjq}, with no additional Riemann terms appearing (except through the new terms that appear in the $a$, $b$ and $c$ tensors which 
define the equations of motion).

We have demonstrated explicitly that our method is successful in evaluating the observable statistics 
of inflationary models with many fields and a curved field-space metric. 
The code we have developed to do this is the second iteration of the  \texttt{PyTransport} package, \texttt{PyTransport\,2.0}, 
and agrees with its predecessor in the case of models which can be written in Euclidean and non-Euclidean 
coordinates (as discussed in \S\,\ref{curved}). Moreover, we have shown that for 
simple 2-field models that the speed of the new code compares with well with that of the canonical model. 
It should be noted, however,  that the the code has not been tested for models  exceeding more than six fields, and that 
we expect time taken to scale poorly with number of fields. Our hope is that this new code will be useful to the 
inflationary cosmology community.

\section*{Acknowledgments}
DJM is supported by a Royal Society University Research Fellowship. JWR acknowledges the support of a studentship 
jointly funded by Queen Mary University 
of London and by the Frederick Perren Fund of the University of London. We thank Karim Malik and 
Pedro Carrilho for helpful discussions and for comments on a previous version of this manuscript.  
We thank David Seery, Sean Butchers, Mafalda 
Dias and Jonathan Frazer for useful discussions related to code development, and David Seery 
for cross checks of the output of our code with that of \texttt{CppTransport}.

\newpage
\begin{appendices}
\section{Initial Conditions}
Here we provide a few more detail of how the initial conditions for the 
transport system are calculated. We recall that $\star$ denotes a time 
long before horizon crossing at which $-\tau \gg 1$, where $\tau$ denotes conformal time, 
and that for a de-Sitter expansion
$\tau = -1/(aH)$.

\subsection{Two point function}
First we consider initial conditions for the two point function for 
the various combinations of covariant field perturbation and momenta correlations. The calculation 
is similar to that presented in Ref.~\cite{Dias:2015rca}, though in that work the time variable 
used for the transport system was e-folds $N$, while in this paper we
use cosmic time, $t$.

\label{apdxA}
\begin{itemize}
  \item Field-Field correlation
\end{itemize}
Beginning with the expression for the two point function of $Q^I$ (\ref{IC2pt}) we consider the $-\tau \gg 1$
limit for the field-field correlations. We find
\begin{equation}
\begin{split}
\langle Q^I (k_1,\tau)Q^J(k_2,\tau)\rangle &= (2\pi)^3\delta(k_1+k_2)\frac{ G^{IJ}}{2k^3}H^2(\tau)(1-ik\tau)(1+ik\tau)\\
&\approx(2\pi)^3\delta(k_1+k_2)\frac{ G^{IJ}}{2k^3}H^2(\tau)\vert k\tau\vert^2\\
&\approx(2\pi)^3\delta(k_1+k_2)\frac{ G^{IJ}}{2a^2k}\,.
\end{split}
\end{equation}
The initial condition for $\Sigma^{IJ}_{*}$ is then
\begin{equation}
\Sigma^{IJ}_{*Re}=\left.\frac{ G^{IJ}}{2a^2k}\right|_*\,, \quad \Sigma^{IJ}_{*Im}=0\,.
\end{equation}

\begin{itemize}
\item Field-Momentum correlation
\end{itemize}
Next recalling that at linear order $P^I = D_{t} Q^I$ and that the covariant derivative of the 
parallel propagator is zero, we consider the leading 
term in the expression for the field-momentum correlation of unequal time 
correlations, and subsequently take equal time limit for the case 
$-\tau \gg 1$. Recalling that $d\tau = dt/a(t)$ we find
\begin{equation}
\begin{split}
\langle Q^I (k_1,\tau_1)P^J(k_2,\tau_2)\rangle &=
(2\pi)^3\delta(k_1+k_2)\frac{ \Pi^{IJ}}{2k^3}H(\tau_1)H(\tau_2)(1+ik\tau_1) \left(\frac{k^2\tau_2}{a}\right)e^{ik(\tau_2-\tau_1)}\\
 & =
(2\pi)^3\delta(k_1+k_2)\frac{ G^{IJ}}{2k^3}H^2(\tau)\left(\frac{k^2\tau}{a}\right)(1-ik\tau)\\
& = (2\pi)^3\delta(k_1+k_2)\left(-\frac{G^{IJ}H}{2ka^2} +i\frac{G^{IJ}}{2a^3} \right).
\end{split}
\end{equation}
The real and imaginary parts of the initial conditions for this case are then
\begin{equation}
\Sigma^{IJ}_{*Re}=\left.-\frac{G^{IJ}H}{2ka^2}\right|_*\,, \quad \Sigma^{IJ}_{*Im}= \left.\frac{G^{IJ}}{2a^3}\right|_*\,.
\end{equation}

\begin{itemize}
\item Momentum-Momentum correlation
\end{itemize}
We follow a similar procedure to consider the momentum-momentum correlation
\begin{equation}
\begin{split}
\langle P^I (k_1,\tau_1)P^J(k_2,\tau_2)\rangle &=
(2\pi)^3\delta(k_1+k_2)\frac{ \Pi^{IJ}}{2k^3}H(\tau_1)H(\tau_2)\left(\frac{ k^2\tau_1}{a}\right)\left( \frac{k^2\tau_2}{a}\right)e^{ik(\tau_2-\tau_1)}\\
 & =
(2\pi)^3\delta(k_1+k_2)\frac{ G^{IJ}}{2k^3}H^2(\tau)\left( \frac{k^4\tau^2}{a^2}\right)\\
& = (2\pi)^3\delta(k_1+k_2)\frac{G^{IJ}k}{2a^4}.
\end{split}
\end{equation}
The initial condition for $\Sigma^{IJ}$ in this case is 
\begin{equation}
\Sigma_{*Re}^{IJ}=\left.\frac{G^{IJ}k}{2a^4}\right|_*\,, \quad \Sigma_{*Im}^{IJ}=0\,.
\end{equation}

\subsection{Three point function}
For the three-point function as discussed in the main text, an integral must be 
evaluated to calculate the initial condition. By substituting Eq.~(\ref{eq:hintabc}) into Eq.~(\ref{intialB}) we obtain the initial condition $B_{*}^{abc}$.
To illustrate how this is evaluated in practice, 
let us consider this explicitly for the case of a field-field-field correlation.
\begin{itemize}
\item a,b,c $\rightarrow$ Field-Field-Field
\end{itemize}
Substituting in the expression for the two-point function we obtain
\begin{equation}
\label{intialBExpanded}
\begin{split}
B_{*}^{abc} =& -\frac{iH^6}{8\Pi_i k_i^3}(1+ik_1\tau)(1+ik_2\tau)(1+ik_3\tau)e^{-ik_s\tau}\times\\ &\int^{\tau_{init}}_{-\infty} \frac{d\eta}{H^2\eta^2}\left[\frac{\dot{\phi}^IG^{JK}}{4H}({\bf{k_2}}\cdot {\bf{k_3}})(1-ik_1\eta)(1-ik_2\eta)(1-ik_3\eta)e^{ik_s\eta}\right. \\
&+ \frac{a^{IJK}_s}{2H^2\eta^2}(1-ik_1\eta)(1-ik_2\eta)(1-ik_3\eta)e^{ik_s\eta}\\
&+\frac{b^{IJK}}{2H^2\eta^2}(1-ik_1\eta)(1-ik_2\eta)k_3^2\eta e^{ik_s\eta}\\
&\left. +\frac{c^{IJK}}{2}k^2_1k^2_2\eta^2(1-ik_3\eta)e^{ik_s\eta} + perms\right] + c.c.\,,
\end{split}
\end{equation}
where we assume that $H$ and $\Pi^{IJ}$ are sufficiently slowly varying 
to be taken as constants and that we can take $\Pi^{IJ} \to G^{IJ}$.

In order to perform the integration we need to know the time dependence of the tensors. As discussed in \S\ref{section4} 
the $a_{IJK}$ tensor contains fast and slow varying parts. The part containing terms quadratic in $\eta$  vary quickly and 
so are included in the integral separately (the first term in Eq.~(\ref{intialBExpanded})), the remaining parts we label $a_s^{IJK}$ and we assume can be considered  constant in time. 
The next step is to perform the integration recalling that the result is dominated by the upper limit (because the integral 
is highly oscillatory into the past). 
Keeping the 
leading and sub-leading terms in $\tau$, and writing in terms of $a$ and $H$, the final result is 
\begin{equation}
\begin{split}
B^{abc}_{*} = & \frac{1}{4a^4}\frac{1}{{{k_1}}\cdot{{k_2}}\cdot{{k_3}}\cdot k_s}\left( - c^{IJK}(k_1,k_2,k_3)\cdot ({{k_1}} \cdot {{k_2}}) - c^{IKJ}(k_1,k_3,k_2)\cdot ({{k_1}} \cdot {{k_3}})
\right.
\\ & 
 - c^{JKI}(k_2,k_3,k_1)\cdot ({{k_2}} \cdot {{k_3}}) +a^2a_s^{IJK} (k_1,k_2,k_3)\\&
  +a^2 a_s^{IKJ}(k_1,k_3,k_2) + a^2 a_s^{JKI}(k_2,k_3,k_1)\\& 
 + a^2 H b^{IJK}(k_1,k_2,k_3)\left(\frac{({{k_1}}+{{k_2}})\cdot {{k_3}}}{{{k_1}}\cdot{{k_2}}} - \frac{K2}{{{k_1}}\cdot{{k_2}}}\right)
 \\& 
+ a^2 H b^{IKJ}(k_1,k_3,k_2)\left(\frac{({{k_1}}+{{k_3}})\cdot {{k_1}}}{{{k_1}}\cdot{{k_3}}} - \frac{K2}{{{k_1}}\cdot{{k_3}}}\right)
 \\& 
+ a^2 H b^{JKI}(k_2,k_3,k_1)\left(\frac{({{k_2}}+{{k_3}})\cdot {{k_1}}}{{{k_2}}\cdot{{k_3}}} - \frac{K2}{{{k_2}}\cdot{{k_3}}}
\right)
\\& \left.\left.+\frac{\dot{\phi}^I}{4H}G^{JK}(-{{k_2}}^2-{{k_3}}^2+{{k_1}}^2)+\frac{\dot{\phi}^J}{4H}G^{IK}(-{{k_1}}^2-{{k_3}}^2+{{k_2}}^2)+\frac{\dot{\phi}^K}{4H}G^{IJ}(-{{k_1}}^2-{{k_2}}^2+{{k_3}}^2) \right)\right|_*\, ,
\end{split}
\end{equation}
where $K2 \equiv k_1 k_2 + k_1 k_3 + k_2 k_3$ and $k_s = k_1+k_2+k_3$.
Repeating for the other correlations we find
\begin{itemize}
  \item  a,b,c $\rightarrow$ Momentum-Field-Field
\end{itemize}
\begin{equation}
\begin{split}
B^{abc}_{*}  = & -\frac{H}{4a^3K3}\left(-\frac{{{k_1^2}}({{k_2}}+{{k_3}})}{k_s} \cdot {{k_1}}\cdot{{k_2}}\cdot{{k_3}}\right)\left( - c^{IJK}(k_1,k_2,k_3)\cdot ({{k_1}}\cdot {{k_2}}) - c^{IKJ}(k_1,k_3,k_2)\cdot ({{k_1}}\cdot {{k_3}} )
\right. 
\\ & \left.
- c^{JKI}(k_2,k_3,k_1)\cdot ({{k_2}}\cdot {{k_3}})+a^2 a_s^{IJK}(k_1,k_2,k_3) + a^2 a_s^{IKJ}(k_1,k_3,k_2) + a^2 a_s^{JKI}(k_2,k_3,k_1) 
\right.
\\ & \left.
+ \frac{G^{JK}\dot{\phi}^I}{4H}(-{{k_2}}^2-{{k_3}}^2+{{k_1}}^2)
+\frac{\dot{\phi}^J}{4H}G^{IK}(-{{k_1}}^2-{{k_3}}^2+{{k_2}}^2)
+\frac{\dot{\phi}^K}{4H}G^{IJ}(-{{k_1}}^2-{{k_2}}^2+{{k_3}}^2) \right)
\\ &
- \frac{H}{4a^3K3}\left(-\frac{{{k_1^2}}\cdot({{k_2}}\cdot {{k_3}})}{k_s}\right)\left( c^{IJK}(k_1,k_2,k_3){{k_1}}^2{{k_2}}^2\left(1+\frac{{{k_3}}}{k_s}\right)+c^{IKJ}(k_1,k_3,k_2){{k_1}}^2{{k_3}}^2\left(1+\frac{{{k_2}}}{k_s}\right) \right.
\\ & \left.
+c^{JKI}(k_2,k_3,k_1){{k_3}}^2{{k_2}}^2\left(1+\frac{{{k_1}}}{k_s}\right) -a^2a_s^{IJK}(k_1,k_2,k_3)\left(K2 -\frac{{{k_1}}\cdot{{k_2}}\cdot{{k_3}}}{k_s} \right) 
\right.
\\ & \left.
- a^2a_s^{IKJ}(k_1,k_3,k_2)\left(K2 -\frac{{{k_1}}\cdot{{k_2}}\cdot{{k_3}}}{k_s} \right)
- a^2a_s^{JKI}(k_2,k_3,k_1)\left(K2 -\frac{{{k_1}}\cdot{{k_2}}\cdot{{k_3}}}{k_s} \right)
\right.
\\ & \left.
+ b^{IJK}(k_1,k_2,k_3)\frac{{{k_1}}\cdot{{k_2}}\cdot{{k_3}}^2}{H} 
+ b^{IKJ}(k_1,k_3,k_2)\frac{{{k_1}}\cdot{{k_3}}\cdot{{k_2}}^2}{H}
+ b^{JKI}(k_2,k_3,k_1)\frac{{{k_2}}\cdot{{k_3}}\cdot{{k_1}}^2}{H} 
\right.
\\ & \left.
-\frac{G^{JK}\dot{\phi}^I}{4H}(-{{k_2}}^2-{{k_3}}^2+{{k_1}}^2)\left(K2+\frac{{{k_1}}\cdot{{k_2}}\cdot{{k_3}}}{k_s}\right)
-\frac{G^{IK}\dot{\phi}^J}{4H}(-{{k_1}}^2-{{k_3}}^2+{{k_2}}^2)\left(K2+\frac{{{k_1}}\cdot{{k_2}}\cdot{{k_3}}}{k_s}\right)\right.
\\ & \left.\left.
-\frac{G^{IJ}\dot{\phi}^K}{4H}(-{{k_1}}^2-{{k_2}}^2+{{k_3}}^2)\left(K2+\frac{{{k_1}}\cdot{{k_2}}\cdot{{k_3}}}{k_s}\right)\right)\right|_*\,,
\end{split}
\end{equation}
where $K3=k_1^3+k_2^3+k_3^3$.
\newpage
\begin{itemize}
  \item  a,b,c $\rightarrow$ Momentum-Momentum-Field
\end{itemize}
\begin{equation}
\begin{split}
B^{abc}_{*} = & - \frac{1}{4a^4K3}\frac{({{k_1}}\cdot{{k_2}}\cdot{{k_3}})^2\cdot {{k_1}}\cdot{{k_2}}}{k_s}\left( -c^{IJK}(k_1,k_2,k_3)\cdot ({{k_1}}\cdot {{k_2}}) - c^{IKJ}(k_1,k_3,k_2)\cdot ({{k_1}}\cdot {{k_3}}) 
\right.
\\ & \left. 
- c^{IJK}(k_2,k_3,k_1)\cdot ({{k_2}}\cdot {{k_3}})
+a^2a_s^{IJK}(k_1,k_2,k_3)+a^2a_s^{IKJ}(k_1,k_3,k_2)+a^2a_s^{JKI}(k_2,k_3,k_1)
\right.
\\ & \left. 
 +a^2Hb^{IJK}(k_1,k_2,k_3)\left(\frac{({{k_1}}+{{k_2}})\cdot {{k_3}} }{{{k_1}}\cdot {{k_2}}} + ({{k_1}}^2 \cdot {{k_2}}^2)\cdot{{k_1}}\cdot{{k_2}}\cdot{{k_3}}^2\right) 
\right.
\\ & \left.
+ a^2Hb^{IKJ}(k_1,k_3,k_2)\left(\frac{({{k_1}}+{{k_3}})\cdot {{k_2}} }{{{k_1}}\cdot {{k_3}}}+ ({{k_1}}^2 \cdot {{k_2}}^2)\cdot{{k_1}}\cdot{{k_3}}\cdot{{k_2}}^2 \right) 
\right.
\\ & \left.
+a^2Hb^{JKI}(k_2,k_3,k_1)\left( \frac{({{k_2}}+{{k_3}})\cdot {{k_1}} }{{{k_2}}\cdot {{k_3}}} +  ({{k_1}}^2 \cdot {{k_2}}^2)\cdot{{k_2}}\cdot{{k_3}}\cdot{{k_1}}^2\right)
\right.
\\ & \left.\left.
-\frac{G^{JK}\dot{\phi}^I}{4H}(-{{k_2}}^2-{{k_3}}^2+{{k_1}}^2)
-\frac{G^{IK}\dot{\phi}^J}{4H}(-{{k_1}}^2-{{k_3}}^2+{{k_2}}^2)
-\frac{G^{IJ}\dot{\phi}^K}{4H}(-{{k_1}}^2-{{k_2}}^2+{{k_3}}^2)
\right)\right|_*\,.
\end{split}
\end{equation}
\begin{itemize}
  \item  a,b,c $\rightarrow$ Momentum-Momentum-Momentum 
\end{itemize}
\begin{equation}
\begin{split}
B^{abc}_{*} =&-\frac{H}{4a^3K3}\frac{{{k_1}}^2{{k_2}}^2{{k_3}}^2}{k_s}\left(c^{IJK}(k_1,k_2,k_3)\cdot ({{k_1}}\cdot {{k_2}})^2\left( 1+\frac{k_3}{k_s}\right) +c^{IKJ}(k_1,k_3,k_2)\cdot ({{k_1}}\cdot {{k_3}})^2\left( 1+\frac{k_2}{k_s}\right)
\right.
\\ & \left.
+ c^{JKI}(k_2,k_3,k_1)\cdot ({{k_3}}\cdot {{k_2}})^2\left( 1+\frac{k_1}{k_s}\right) -a^2a_s^{IJK}(k_1,k_2,k_3)\left(K2-\frac{{{k_1}}\cdot{{k_2}}\cdot{{k_3}}}{k_s}\right)
\right.
\\ & \left.
-a^2a_s^{IKJ}(k_1,k_3,k_2)\left(K2-\frac{{{k_1}}\cdot{{k_2}}\cdot{{k_3}}}{k_s}\right)
-a^2a_s^{JKI}(k_2,k_3,k_1)\left(K2-\frac{{{k_1}}\cdot{{k_2}}\cdot{{k_3}}}{k_s}\right)
\right.
\\ & \left.
+\frac{b^{IJK}(k_1,k_2,k_3)}{H}{{k_1}}{{k_2}}\cdot{{k_3}}^2
+\frac{b^{IKJ}(k_1,k_3,k_2)}{H}{{k_1}}\cdot{{k_3}}\cdot{{k_2}}^2
+\frac{b^{IJK}(k_2,k_3,k_1)}{H}{{k_2}}\cdot{{k_3}}\cdot{{k_1}}^2
\right.
\\ & \left.
-\frac{G^{JK}\dot{\phi}^I}{4H}(-{{k_2}}^2-{{k_3}}^2+{{k_1}}^2)\left(K2+\frac{{{k_1}}\cdot{{k_2}}\cdot{{k_3}}}{k_s}\right)
-\frac{G^{IK}\dot{\phi}^J}{4H}(-{{k_1}}^2-{{k_3}}^2+{{k_2}}^2)\left(K2+\frac{{{k_1}}\cdot{{k_2}}\cdot{{k_3}}}{k_s}\right)
\right.
\\ & \left.\left.
-\frac{G^{IJ}\dot{\phi}^K}{4H}(-{{k_1}}^2-{{k_2}}^2+{{k_3}}^2)\left(K2+\frac{{{k_1}}\cdot{{k_2}}\cdot{{k_3}}}{k_s}\right)
\right)\right|_*\,.
\end{split}
\end{equation}
\end{appendices}
\bibliographystyle{JHEP}
\bibliography{mybib}

\end{document}